\documentclass[showpacs,twocolumn,preprintnumbers,superscriptaddress,prb,amssymb,floatfix]{revtex4-1}
\usepackage{graphicx}
\usepackage{dcolumn}
\usepackage{bm}
\usepackage{amsmath}
\usepackage{amsfonts}
\usepackage{amssymb}
\usepackage{ulem}
\usepackage{color}

\begin{document}
\title{Transport properties of 2D graphene containing structural defects}

\author{Aur\'elien Lherbier}
\affiliation{Universit\'e catholique de Louvain (UCL), Institute of Condensed Matter and Nanoscience (IMCN), Chemin des \'etoiles 8 (ETSF-NAPS, SC17, Bte L7.03.01), 1348 Louvain-la-Neuve, Belgium}
\author{Simon M.-M. Dubois}
\affiliation{University of Cambridge, Cavendish Laboratory, Theory of Condensed Matter group, JJ Thomson Avenue, Cambridge CB3 0HE, United-Kingdom}
\author{Xavier Declerck}
\affiliation{Universit\'e catholique de Louvain (UCL), Institute of Condensed Matter and Nanoscience (IMCN), Chemin des \'etoiles 8 (ETSF-NAPS, SC17, Bte L7.03.01), 1348 Louvain-la-Neuve, Belgium}
\author{Yann-Michel Niquet}
\affiliation{CEA-UJF, INAC, SP2M/L\_Sim, 17 rue des Martyrs, 38054 Grenoble Cedex 9, France}
\author{Stephan Roche}
\affiliation{Institut Catal\`{a} de Nanotecnologia (ICN) and CIN2, UAB Campus, E-08193 Barcelona, Spain}
\affiliation{Instituci\'o Catalana de Recerca i Estudis Avan{\c c}ats (ICREA), 08010 Barcelona, Spain}
\author{Jean-Christophe Charlier}
\affiliation{Universit\'e catholique de Louvain (UCL), Institute of Condensed Matter and Nanoscience (IMCN), Chemin des \'etoiles 8 (ETSF-NAPS, SC17, Bte L7.03.01), 1348 Louvain-la-Neuve, Belgium}
\date{\today}

\begin{abstract}
We propose an extensive report on the simulation of electronic transport in 2D graphene in presence of structural defects. Amongst the large variety of such defects in ${\rm\textit{sp}}^{2}$ carbon-based materials, we focus on the Stone-Wales defect and on two divacancy-type reconstructed defects. First, based on ab initio calculations, a tight-binding model is derived to describe the electronic structure of these defects. Then, semiclassical transport properties including the elastic mean free paths, mobilities and conductivities are computed using an order-N real-space Kubo-Greenwood method. A \textit{plateau} of minimum conductivity ($\sigma_{sc}^{min}=4e^2/\pi h$) is progressively observed as the density of defects increases. This saturation of the decay of conductivity to $\sigma_{sc}^{min}$ is associated with defect-dependent resonant energies. Finally, localization phenomena are captured beyond the semiclassical regime. An Anderson transition is predicted with localization lengths of the order of tens of nanometers for defect densities around $1\%$.
\end{abstract}

\pacs{73.23.-b, 72.15.Rn, 72.80.Vp, 71.23.An}

\maketitle

\section{Introduction}

\begin{figure*}[ht!]
\begin{center}
\leavevmode
\includegraphics[width=1.0\textwidth]{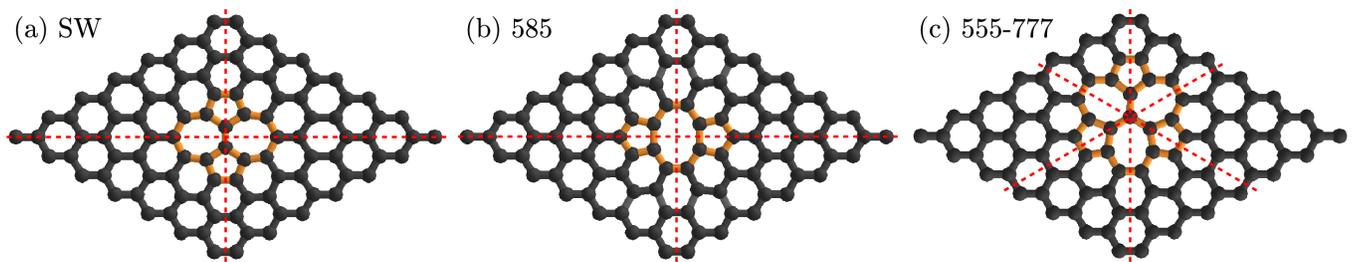}
\caption{(color online). Schematic of the three structural defects: (a) Stone-Wales, (b) 585 and (c) 555-777 divacancies. Symmetry axis are drawn in red dashed lines.}
\label{fig1}
\end{center}
\end{figure*}

Most exceptional properties of graphene are closely related to its crystalline structure and its reduced dimensionality. The combination of a honeycomb lattice and a real two dimensional material is at the origin of the linear dispersion and pseudospin symmetry of graphene low energy carriers. This gives rise to a wealth of interesting properties such massless Dirac fermions~\cite{NovoselovRMP}, reduced back scattering, Berry phase~\cite{Zhang} and weak antilocalization~\cite{FOrtman}.\\
Besides, some transport characteristics of graphene, such as high carrier mobilities~\cite{Bolotin} and long electronic coherence lengths~\cite{Miao}, are highly interesting from a technology point of view. Graphene could also match many request of the process and design of nanoelectronic devices. Indeed, because of its 2D geometry, strategies of top-down fabrication approach such as photolithography, intensively used in the mass production industry, could be more easily engineered and applied to design graphene-based nanodevices architectures. Mainly for these reasons, graphene is expected to become a material of choice for future nanoelectronic.\\
Yet, the lack of electronic (or transport) gap in pristine graphene is an issue that has to be overcome for achieving standard electronic devices such as field effect transistors (FET). As it has been the case for silicon, the tunning of graphene electronics is an essential step towards applications. For this reason, controlled defect engineering in ${\rm\textit{sp}}^{2}$ carbon-based materials has become a topic of great excitement~\cite{Krasheninnikov}. Indeed, the electronic (and transport) properties of carbon nanotubes~\cite{Charlier} and graphene-based materials~\cite{Suenaga,Lusk,Cresti} can be considerably enriched by chemical modifications, including substitution and molecular doping~\cite{lherbier_dopant,Wang} as well as functionalization. Another approach to tune graphene's properties is provided by ion or electron beam irradiation which introduces structural defects (\textit{e.g.}\ vacancies) in ${\rm\textit{sp}}^{2}$ carbon-based nanostructures. As an example, convincing room-temperature signatures of an Anderson regime in ${\rm Ar}^{+}$ irradiated carbon nanotubes have been reported~\cite{Biel1,Biel2}. In contrast, the conductivity of irradiated two-dimensional graphene saturates at the Dirac point above $e^{2}/h$ even down to cryogenic temperatures~\cite{JHChen2009}, suggesting a stronger robustness of defective graphene. Nonetheless, many symmetries are broken when lattice disorder, such as structural defects, is introduced into a perfect honeycomb lattice~\cite{Libisch}. Accordingly, structural defects in such materials should strongly affect the transport properties. We will show that a strong (Anderson) localization regime could be observed but, only in regions of energy corresponding to specific resonances directly associated to a certain type of defect and which not necessarily coincide with the Dirac point. Another reason why irradiated two-dimensional graphene seems to exhibit strong robustness compared to irradiated carbon nanotubes for an equivalent amount of structural defects, is the dimensionality. Indeed, the scaling theory of localization predicts a different behavior of the localization phenomena in one and two dimensional materials~\cite{Lherbier08}.\\
The paper is elaborated as follow: In section \ref{sec1}, we describe the geometry and the electronic properties of structural defects with ab initio and tight-binding techniques. Then, in section \ref{sec2}, we present the real-space Kubo-Greenwood methodology and discuss the transport properties of large graphene planes with a realistic structural disorder. We also perform an in-depth analysis of the localization phenomena obtained within the present approach. Finally, in section \ref{sec3}, we conclude on our results and discuss about the possibility to validate our theoretical predictions with experiments.

\section{Electronic properties of structural defects}\label{sec1}

\subsection{Geometry of structural defects}

Structural defects exist in various forms in graphene~\cite{Banhart,Kotakoski,CockaynePRB2011,Cockayne} and more generally in ${\rm\textit{sp}}^{2}$ carbon-based materials. In this paper, three types of structural defects are considered, the Stone-Wales (SW) defect and two possible reconstructions of the divacancy. These three defects are non-magnetic, in contrast with monovacancies for instance which contain an unbounded carbon atom. Consequently, the issue of induced magnetization and spin interaction between defects does not need to be taken into account. The SW defect is a  well known and common defect in ${\rm\textit{sp}}^{2}$ carbon-based materials~\cite{Stone-Wales-paper} which consists in a $90$ degree rotation of a carbon-carbon bond. This topological transformation yields to the formation of two heptagons connected with two pentagons (Fig.\ref{fig1}.a). Ma \textit{et al.} have reported in a theoretical study~\cite{Ma} that in graphene a SW defect could produce a slight out-of-plane deformation. In this situation, the carbon bonds in the close neighboring of the SW defect are no more fully ${\rm\textit{sp}}^{2}$ hybridized but become partially ${\rm\textit{sp}}^{3}$-like hybridized. Considering this theoretical prediction would imply that the new orbitals hybridization could induce spin-orbit coupling as it has been predicted for the case of ripples~\cite{Huertas-Hernando}. However, such out-of-plane deformations have never been obtained in our calculations. To ensure that the in-plane geometry was not a metastable state related to a local minimum of total energy, additional calculations have been performed with a starting geometry containing out-of-plane atoms. After structural relaxation, the out-of-plane atoms always go back to the plane. Moreover, to our knowledge no experimental STM measurements have confirmed this behavior predicted in Ref~\cite{Ma}. Therefore, only the in-plane geometry has been considered in this article.\\
Vacancies are missing carbon atoms in the honeycomb lattice. They can not be considered as a reversible geometrical modification of the ideal graphene plane and thus strictly speaking, they do not belong to the class of topological defects. Vacancies can be created by irradiating the graphene plane with ions such as ${\rm Ar}^{+}$ for instance. Single vacancies (or monovacancies) migrate easily in the graphene plane and are stabilized when they recombine with another one to give a divacancy defect~\cite{Krasheninnikov,GundoLee,Kim}. Two kinds of divacancy defects are explored here. In the first case, the reconstruction yields to the formation of 2 pentagons and 1 octagon (so-called 585, Fig.\ref{fig1}.b), while in the second case it yields to the formation of 3 pentagons and 3 heptagons (so-called 555-777, Fig.\ref{fig1}.c). According to our ab initio calculations the formation energy of the 555-777 divacancy is smaller than the one of the 585 divacancy by about $0.9$eV. This stabilization of the 555-777 divacancy with regards to the 585 divacancy in graphene, which contrasts to the case of carbon nanotubes, is also reported by G.-D. Lee {\it et al.}~\cite{GundoLee}. A third kind of divacancy, not studied here, has also been reported~\cite{Banhart,Meyer}. This third type of divacancy has a larger extension and is thus more complicated to parametrize within a tight-binding model. Actually, it involves 9 carbon rings including  4 pentagons, 1 hexagon and 4 heptagons (so-called 5555-6-7777).
To conclude on the geometry analysis of these defects, one notes that the 585 divacancy as well as the SW defect possess a $D_{2h}$ symmetry since two orthogonal symmetry axis can be defined, whereas the 555-777 divacancy possess a $D_{3h}$ symmetry (see Fig.\ref{fig1}). The presence of these three structural defects has already been experimentally reported in graphene by means of STM experiments~\cite{Suenaga,Ugeda} or TEM images~\cite{Meyer}, and their influence on the transport properties deserves in-depth inspection.

\subsection{Ab initio simulations}\label{sectionabinitio}

\begin{figure*}[ht!]
\begin{center}
\leavevmode
\includegraphics[width=1.0\textwidth]{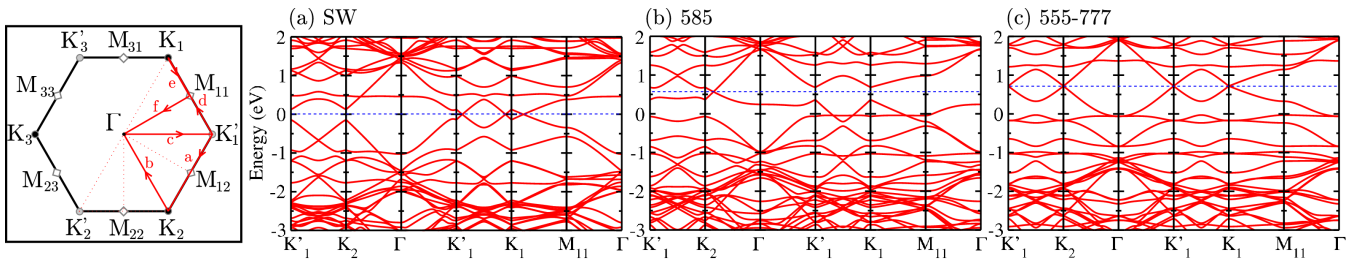}
\caption{(color online). Electronic band structures computed using the SIESTA package along high-symmetry lines in the Brillouin zone (left panel) for $7\times7$ supercell containing, (a) one SW defect, (b) one 585 divacancy, (c) one 555-777 divacancy. The Fermi energy is set to zero and the Dirac energy is indicated with a horizontal blue dashed line.}
\label{fig2}
\end{center}
\end{figure*}

In order to study these three defects, ab initio simulations are first employed. The calculations have been performed using the SIESTA package~\cite{SIESTA} in the generalized gradient approximation (GGA) for the exchange-correlation functional in the Ceperley-Alder form~\cite{Ceperley-Alder}, as parametrized by Perdew and Zunger~\cite{Perdew-Zunger}. Troullier-Martins pseudopotentials are used to account for the core electrons~\cite{Troullier-Martins}. The valence electron wave functions are expanded in a double-$\zeta$ polarized basis set of finite-range numerical pseudoatomic orbitals~\cite{Artacho}.\\
The ab initio calculations have two objectives. On one hand, it allows to obtain the detailed geometry of the defect, and on the other hand, it allows to elaborate a tight-binding model for an isolated defect. In both cases, a convergence study with regards to the size of the cell used in calculations needs to be performed. The supercell technique, which consists in placing the defect in a cell which is a multiple of the standard unitcell of the pristine system, is therefore employed.\\
First, a structural relaxation of the supercell containing one defect is performed for different supercell sizes, i.e. $4\times4$, $5\times5$, $6\times6$ and $7\times7$ ($1\times1$ being the conventional unitcell). To ensure that the defect does not interact with its repeated images, the potential associated to the defect must tend to zero at the edges of the supercell. The $7\times7$ supercells are found to be large enough for each defect to obtain such a convergence. Final positions of the atoms in the vicinity of the defects are then extracted and will be used to construct the tight-binding models. The $7\times7$ supercells after structural relaxation of each defect are displayed in Fig.\ref{fig1}. The SW defect does not induce strong distortions of the neighboring hexagon rings, whereas the 585 divacancy exhibits a more pronounced impact. Indeed in the latter case, a distortion of the hexagons rings along the vertical axis (aligned with octagon) is observed. A slight distortion is also observed for the 555-777 divacancy but smaller than for the 585 one.\\
In Fig.\ref{fig2}, the electronic properties of the defects are examined through the study of their respective band structures. The electronic bands are plotted along an extended path connecting high symmetry points in the Brillouin zone. The path used here (see left panel of Fig.\ref{fig2}), which is longer than the usual K-$\Gamma$-M-K path used for graphene, has been carefully chosen to evidence all useful details. Indeed, as the structural defects break the symmetry of the pristine graphene lattice, the hexagonal symmetry of the Brillouin zone is also lifted. As a consequence, paths in the Brillouin zone that were equivalent in the pristine case can become inequivalent when such a type of defect is introduced. By definition, if there are no other symmetries apart the time-reversal symmetry, half of the Brillouin zone needs to be examined which implies at least $10$ branches (full and dotted red lines in left panel of Fig.\ref{fig2}). Here, a minimal continuous path composed of $6$ branches have been chosen in order to depict all important details for each defect. This extended path is labeled \textit{a-b-c-d-e-f} in the left panel of Fig.\ref{fig2}.\\
Then, for sake of comparison between band structures of the three defects, the Fermi energy has been set to zero and the energy associated with the Dirac point has been indicated by a horizontal blue dashed line. The Dirac cone band crossing occurs at an energy $E=0.7$eV above the Fermi energy in case of the 555-777 divacancy (Fig.\ref{fig2}.c), thus suggesting that this defect has a p-type doping character since the Fermi energy has been shifted below the Dirac point energy, whereas both energies were aligned in the pristine case. Obviously, the magnitude of the shift of the Fermi energy is related to the concentration of defects ($n_d$). One defect in a $7\times7$ supercell corresponds to $n_d\sim3.9\,\,10^{13}$cm$^{-2}$ or equivalently $n_d\sim1$\% which is the maximum concentration that will be considered in section~\ref{sec2} for transport properties. For the 585 divacancy (Fig.\ref{fig2}.b), the Dirac point energy is located at $E=0.6$eV above the Fermi energy, also implying a p-type doping character although slightly weaker than for the 555-777 divacancy. Finally the Dirac point energy and the Fermi energy are both aligned in case of a SW defect (Fig.\ref{fig2}.a) which indicates a no doping effect.\\
As mentioned in the previous paragraph, the 555-777 divacancy possesses a $D_{3h}$ symmetry which is the same symmetry group as a single Dirac cone (at K or at K$^{\rm{'}}$). Indeed, even if very close to the Dirac point energy the Dirac cone is isotropic, at higher energies it has been demonstrated theoretically~\cite{McCann,Mucha} and by ARPES measurements~\cite{Zhou} that a trigonal warping emerges. This effect is a signature of the $D_{3h}$ symmetry of a single Dirac cone related to the existence of two inequivalent triangular sublattices. Since the 555-777 divacancy has the same symmetry, the band structure of the defected system preserves the initial symmetry present in pristine graphene. Therefore, the usual electronic path composed of only three branches should be sufficient. It is actually easy to see for instance that in Fig.\ref{fig2}.c, paths $\Gamma$-K$_2$ and $\Gamma$-K$_1^{\rm{'}}$ are equivalent as well as paths K$_1^{\rm{'}}$-K$_2$ and K$_1^{\rm{'}}$-K$_1$. On the contrary, the 585 divacancy and the SW defect possess a $D_{2h}$ symmetry which breaks the symmetry of the Dirac cone and as a consequence the latter is shifted. For the 585 divacancy, the first Dirac cone band crossing corresponding to K valley is shifted along the direction K$_2\rightarrow\Gamma$, whereas the second Dirac cone band crossing corresponding to K$^{\rm{'}}$ valley is shifted in the opposite direction (K$_3^{\rm{'}}\rightarrow\Gamma$). The latter is not visible in Fig.\ref{fig2}.b since the path K$_3^{\rm{'}}$-$\Gamma$ is not represented (but is the symmetric of K$_2$-$\Gamma$). For the SW defect the shift is reversed, i.e. the first Dirac cone band crossing corresponding to K valley is shifted along the direction $\Gamma\rightarrow\rm{K}_2$ (or equivalently along K$_1\rightarrow\rm{K}_1^{\rm{'}}$), whereas the second Dirac cone band crossing corresponding to K$^{\rm{'}}$ valley is shifted along the opposite direction (K$_1^{\rm{'}}\rightarrow\rm{K}_1$). For both defects, paths $\Gamma$-K$_2$ and $\Gamma$-K$_1^{\rm{'}}$ are now inequivalent as well as paths K$_1^{\rm{'}}$-K$_2$ and K$_1^{\rm{'}}$-K$_1$. This shift of the Dirac cone induced by symmetry breaking has already been reported in case of uniaxial strain~\cite{Castro-Neto_RMP}, but to our knowledge not in case of point defects.\\
Some features of the transport properties could already be anticipated from these electronic band structures analysis. For instance, rather flat bands close to the Dirac point energy are observed. These flat bands have to be related to the presence of the defect, and can be seen as resonance energies associated with electrons localized around the defect. Such localized states are in general responsible of reduced transport properties. For the SW defect, a flat band is observed around $E=0.5$eV above the Dirac point energy. A second flat band is located at $E=-1.75$eV but is embedded in many dispersive bands. For the 585 divacancy, two regions of flat bands are observed around $0$eV and $0.25$eV corresponding respectively to energies $E\sim-0.6$eV and $E\sim-0.35$eV below the Dirac point energy. Another flat band is also observed around $-1$eV but again many other dispersive bands stands at this energy. Finally, for the 555-777 divacancy, much more flat band regions are observed but far away from the Dirac point energy. Indeed, flat bands are observed around $1.35$eV, $0$eV and $-1$eV which corresponds respectively to energy $E=0.45$eV above the Dirac point energy, $E=-0.7$eV and $E=-1.7$eV below the Dirac point energy.\\
\begin{figure}[ht!]
\begin{center}
\leavevmode
\includegraphics[width=1.0\columnwidth]{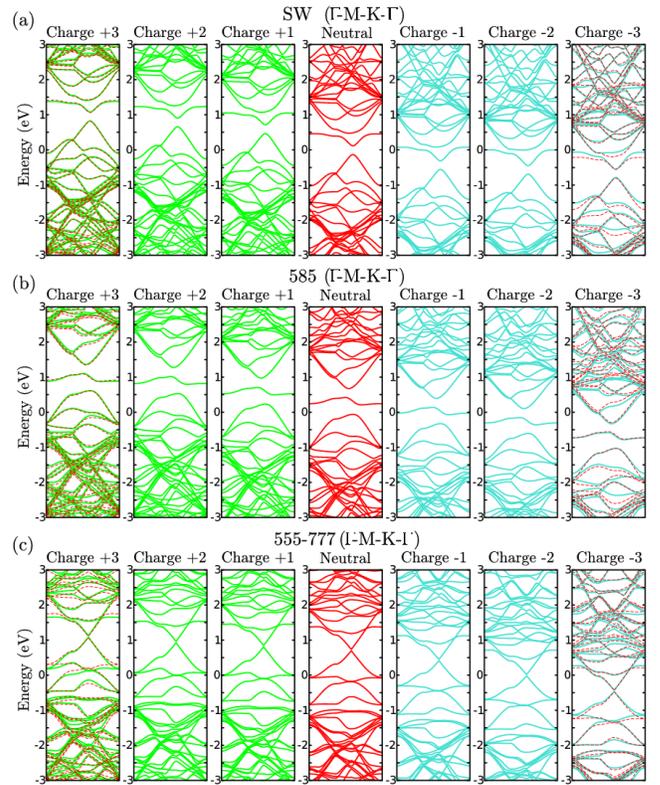}
\caption{(color online). Electronic band structures computed using the SIESTA package along high-symmetry lines $\Gamma$-M-K-$\Gamma$ for $7\times7$ supercell containing, (a) one SW defect, (b) one 585 divacancy, (c) one 555-777 divacancy, and with different excess charge ($\left\lbrace-3,-2,-1,+1,+2,+3\right\rbrace\,\vert e \vert$). The Fermi energy is set to zero.}
\label{fig3}
\end{center}
\end{figure}
Finally, one concludes the analysis of band structures by investigating the case of charged systems. This is reported here as a validation of the rigid band model that will be used in the next section in order to discuss transport properties of defective graphene at different energies. In contrast with the experimental configuration where the applied gate voltage effectively fill or deplete the active part of the device with electrons, our transport model relies on the rigid shift of the Fermi energy in order to describe out-of-equilibrium situations. Indeed, the tight-binding and transport models presented in the next paragraphs are essentially a model of the equilibrium case and can not account for the detailed impact of the excess of charge. In Fig.\ref{fig3}, the validity of this rigid band approximation, which neglects these charge effects, is verified by looking at the impact of net charges on the electronic band structures.
Calculations have been performed for systems with excess charge of $\left\lbrace-3,-2,-1,+1,+2,+3\right\rbrace\,\vert e \vert$ where $\vert e \vert$ is the elementary charge. Note that, computationally, the added charges are compensated by a background jellium. For a $7\times7$ supercell, a charge of $3\,\vert e \vert$ corresponds to an carrier density of $n\sim1.2\,\,10^{14}$cm$^{-2}$. Fig.\ref{fig3} shows that the overall band structure is not strongly modified by the added charges up to $\pm3\,\vert e \vert$ especially in the region $\left[-1,+1\right]$eV around the Fermi energy. For higher energies, the approximation of rigid defect model is more questionable. Also, it is interesting to note that, when the conventional path composed of $3$ branches ($\Gamma$-K-M-$\Gamma$) in the Brillouin zone is used to plot the band structure (see Fig.\ref{fig3}), a fictitious energy gap is observed. This pinpoints the importance of the symmetry breaking of the Brillouin zone and the need of such an extended path as used in Fig.\ref{fig2}.

\subsection{Tight-binding models}

\begin{figure}[ht!]
\begin{center}
\leavevmode
\includegraphics[width=1.0\columnwidth]{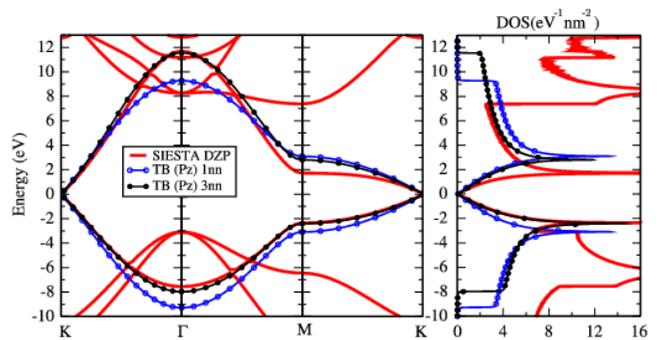}
\caption{(color online). Electronic band structures and corresponding density of states (DOS) computed using the SIESTA package with a double-$\zeta$ polarized (DZP) basis set (red lines) along K-$\Gamma$-M-K path for $1\times1$ supercell (unitcell). The TB band structures for a $1^{\rm{rst}}$ nearest neighbors model ($1$nn, blue lines with open square symbols) and for a $3^{\rm{rd}}$ nearest neighbors model ($3$nn, black lines with filled circle symbols) are also plotted. The Fermi energy is set to zero.}
\label{fig4}
\end{center}
\end{figure}

The tight-binding (TB) models for pristine graphene and for the three defects are derived from the previous ab initio calculations. A common way to obtain the TB parameters is to choose a set of points $E(k)$ in the ab initio (or experimental) band structure and then to use them as constraints in a fit procedure (based on gene algorithm for instance). The TB parameters are adjusted by the fit to reproduce the full band structure as well as possible. Here, a different strategy is applied. Since the SIESTA Hamiltonian is expressed in a localized orbitals basis set, the TB parameters can be directly \textit{extracted} by performing successive operations on this SIESTA Hamiltonian. In particular, the basis set has to be reduced to a single ${\rm\textit{p}_z}$ orbitals orthogonal basis. The details of the technique will be presented in a separated paper~\cite{Siesta-to-TB}.\\
Most TB studies use a $1^{\rm{rst}}$ nearest neighbors $\pi$-$\pi^{*}$ TB model to describe graphene electronic properties. However, in the present work, a two centers $3^{\rm{rd}}$ nearest neighbors $\pi$-$\pi^{*}$ orthogonal TB model has been chosen. Indeed, already for pristine graphene, a $3^{\rm{rd}}$ nearest neighbors $\pi$-$\pi^{*}$ model gives much more reasonable results since it allows in particular to recover the existing asymmetry between valence ($\pi$) and conduction ($\pi^{*}$) bands. In contrast, a $1^{\rm{rst}}$ nearest neighbors $\pi$-$\pi^{*}$ model produces a totally symmetric band structure as shown in Fig.\ref{fig4}. The comparison between the ab initio and the $3^{\rm{rd}}$ nearest neighbors model band structures and total densities of states (DOS) is very satisfactory. However, the K-M branch in the conduction band side remains not exactly reproduced by this $3^{\rm{rd}}$ nearest neighbors model. The TB parameters of this model are composed of a single on-site term $\varepsilon_{{\rm\textit{p}}_{z}}$ and three hopping terms $\gamma_0^1$, $\gamma_0^2$ and $\gamma_0^3$ corresponding respectively to $1^{\rm{rst}}$,$2^{\rm{nd}}$ and $3^{\rm{rd}}$ nearest neighbors. The pristine graphene Hamiltonian then reads as
\begin{eqnarray}
&&H=\sum_{i} \left\langle \phi_i \vert \varepsilon_{{\rm\textit{p}}_{z}} \vert \phi_i \right\rangle  + \nonumber \\
 &&\sum_{i,<j,k,l>} \left( \left\langle \phi_i \vert \gamma_{0}^{1} \vert  \phi_j \right\rangle + \left\langle \phi_i \vert \gamma_{0}^{2} \vert  \phi_k \right\rangle+ \left\langle \phi_i \vert \gamma_{0}^{3} \vert  \phi_l \right\rangle \right) 
\end{eqnarray}
with $\varepsilon_{{\rm\textit{p}}_{z}}=0.59745$eV and $\gamma_{0}^{1}=-3.09330$eV, $\gamma_{0}^{2}=0.19915$eV, $\gamma_{0}^{3}=-0.16214$eV. The sum on index $i$ run over all carbon ${\rm\textit{p}_z}$ orbitals. The sums over $j,k,l$ indexes run over all ${\rm\textit{p}_z}$ orbitals corresponding respectively to $1^{\rm{rst}}$,$2^{\rm{nd}}$ and $3^{\rm{rd}}$ nearest neighbors of the $i^{\rm{th}}$ ${\rm\textit{p}_z}$ orbital.

\begin{figure}[ht!]
\begin{center}
\leavevmode
\includegraphics[width=1.0\columnwidth]{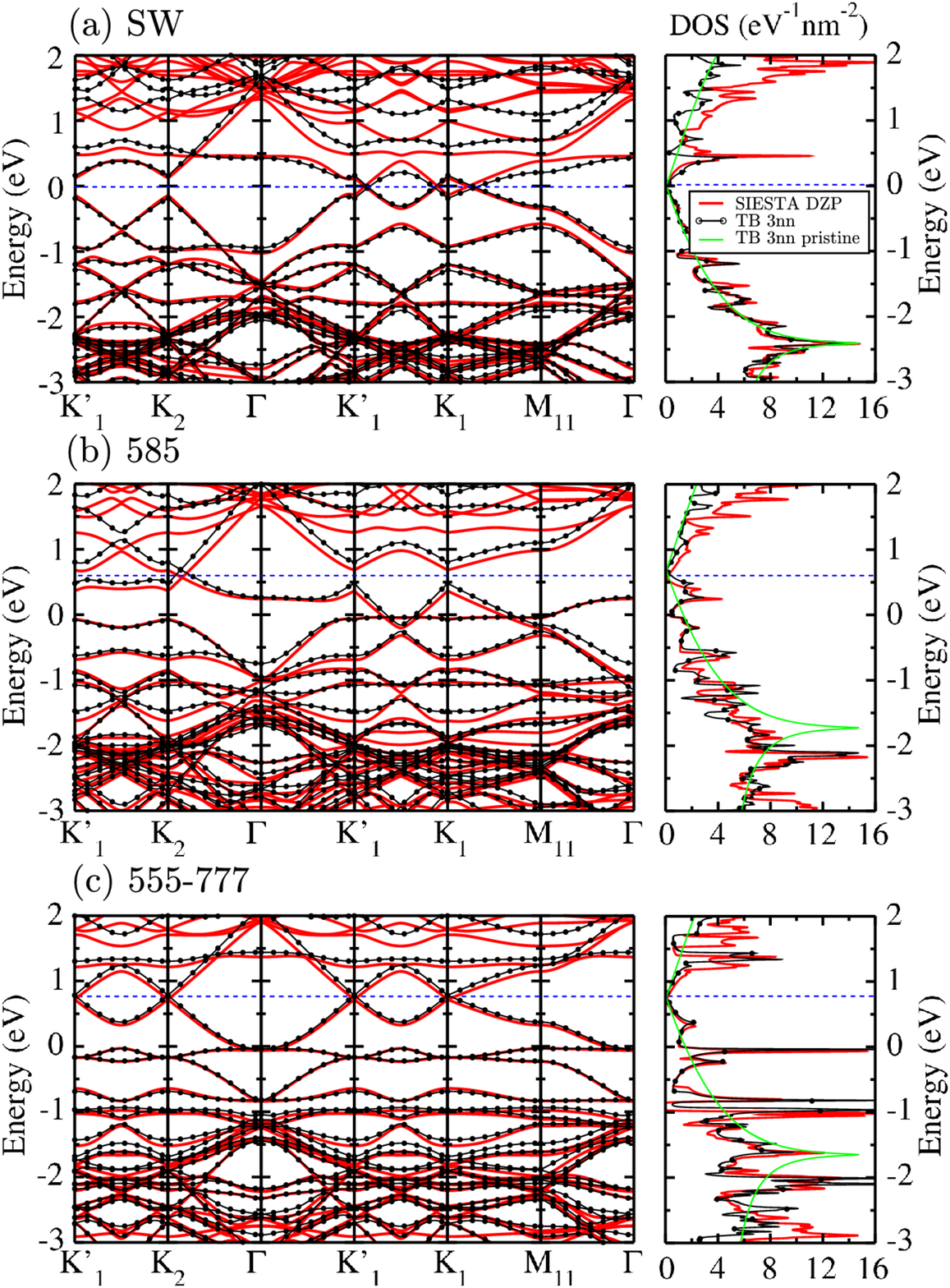}
\caption{(color online). Ab initio electronic band structure (red lines) computed using SIESTA with a double-$\zeta$ polarized (DZP) basis set along high-symmetry lines in the Brillouin zone (see left panel of Fig.\ref{fig2}) for $7\times7$ supercell containing, (a) one SW defect, (b) one 585 divacancy, (c) one 555-777 divacancy. The TB band structures for a $3^{\rm{rd}}$ nearest neighbors model ($3$nn, black lines with filled circle symbols) is also plotted. The Fermi energy is set to zero and the energy of the Dirac point is indicated with a horizontal blue dashed line. In right panels, the corresponding DOS are plotted. The pristine graphene DOS for a $3^{\rm{rd}}$ nearest neighbors model is superimposed for sake of comparison (green thin lines).}
\label{fig5}
\end{center}
\end{figure}

To obtain the local TB parameters corresponding to the defect potential, the same extraction technique as for the pristine graphene have been used. In the defect potential, only on-site modifications have been considered, but the new arrangement of neighbors for carbon atoms in the core of the defects have also been carefully taken into account. In case of SW defect for instance, the rotation of the carbon-carbon bond yields to a modification of first, second and third nearest neighbors for carbon atoms in the vicinity of the rotated bond.
To check the validity of the TB parametrization of the defects, the ab initio and the TB band structures have been compared for a $7\times7$ supercell containing one defect. In Fig.\ref{fig5}, the TB band structure (black dotted lines) is superimposed with the ab initio band structure (red full lines) already reported in Fig.\ref{fig2}. A good agreement is obtained especially for the valence bands. The conduction band side seems to be less accurate but this is uniquely due to the inability of pristine graphene TB model to reproduce conduction band along K-M branch, as already illustrated in Fig.\ref{fig4}. As a consequence, some of the conduction bands are shifted to higher energies. A comparison of ab initio and TB DOS is also provided in the right panels of Fig.\ref{fig5}.

\section{Transport properties}\label{sec2}

\begin{figure}[ht!]
\begin{center}
\leavevmode
\includegraphics[width=1.0\columnwidth]{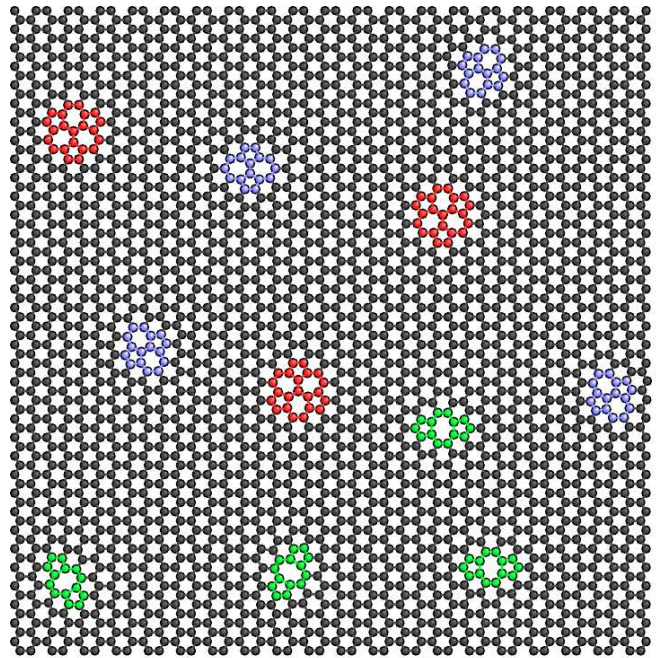}
\caption{(color online). Example of a small piece of defected graphene plane generated using the TB models developed for the three defects. Both position and orientation of the defects are chosen randomly.}
\label{fig6}
\end{center}
\end{figure}

In this section, the transport properties of large defective graphene planes is investigated. Although the electronic structure study performed in the prior section already allows to draw some preliminary conclusions on the impact of structural defects on transport, the randomness character of a realistic disorder at the mesoscopic scale has to be taken into account. Therefore, based on the developed TB models for isolated defects, large graphene planes containing randomly distributed defects of various nature and density are built (Fig.\ref{fig6}). It is important to pay attention to the various possible orientations of the defects. As explained previously, the structural defects possess specific symmetry groups smaller than the one of the honeycomb lattice. Thereby, for the SW and the 585 divacancy there exist three possible orientations, whereas there are only two for the 555-777 divacancy. If a given orientation is overrepresented, anisotropic transport properties would be observed. In our disorder model the defects are randomly oriented to avoid this effect. The case of graphene planes containing a \textit{mixture} of defects is also studied in order to be as close as possible to the experimental situation. In a previous paper, the case of graphene planes containing only one type of defect but with different densities has been investigated~\cite{PRLKubo2011}. For a mixture of several types of defects, the results can be almost derived by linear superposition of transport properties obtained for one specific type of defects as it will be demonstrated in the next paragraph.

\subsection{Density of states}

\begin{figure*}[ht!]
\begin{center}
\leavevmode
\includegraphics[width=1.0\textwidth]{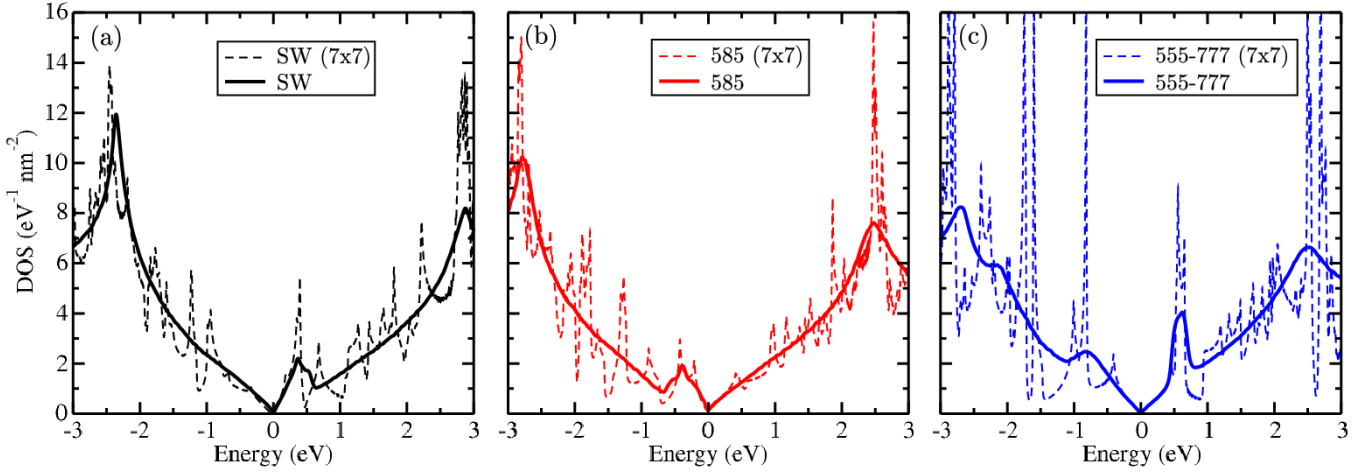}
\caption{(color online). TB DOS for a $7\times7$ supercell (dashed lines) containing, (a) one SW defect, (b) one 585 divacancy, (c) one 555-777 divacancy, i.e. $n_d\sim 1$\%, and for a large plane (thick lines) with $n_d=1$\% of, (a) SW defects, (b) 585 divacancies, (c) 555-777 divacancies, distributed and oriented randomly.}
\label{fig7}
\end{center}
\end{figure*}
\begin{figure*}[ht!]
\begin{center}
\leavevmode
\includegraphics[width=1.0\textwidth]{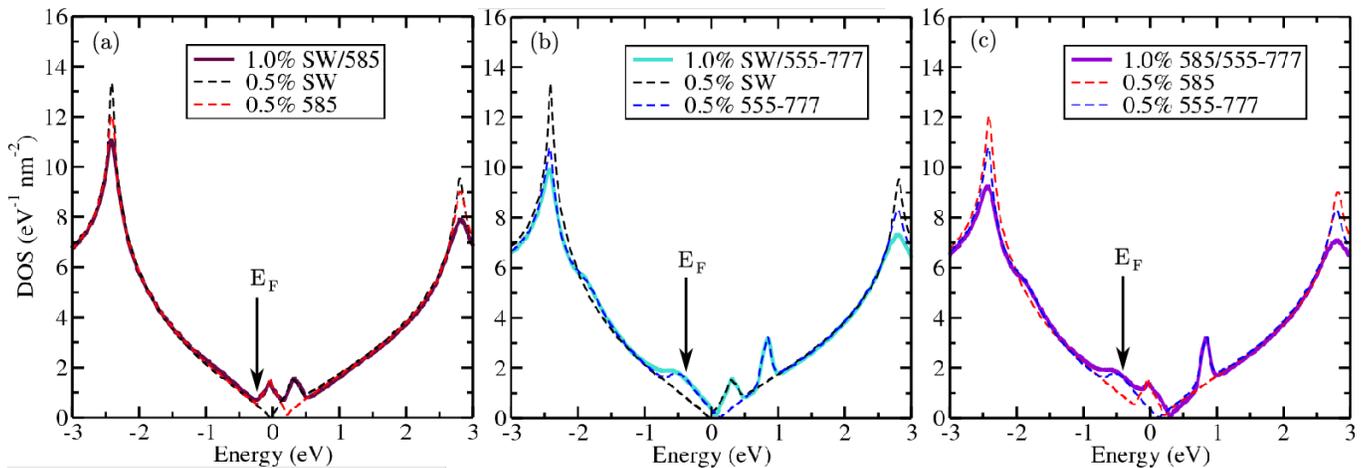}
\caption{(color online). TB DOS for a large graphene plane with $n_d=1$\% defects (thick solid lines) of, (a) SW/585, (b) SW/555-777, and (c) 585/555-777. These DOS are compared with DOS obtained with $n_d=0.5$\% of each defects separately (dashed lines). The position of the Fermi energy is indicated by a vertical arrow.}
\label{fig8}
\end{center}
\end{figure*}

Before achieving an in-depth analysis of the transport properties, it is instructive to examine the density of states of randomly disordered graphene planes.
Indeed, a first guidance to understand the transport features can be obtained by investigating the band structures and the DOS of the system as it has been performed in paragraph \ref{sectionabinitio} for defects in small supercells. Here, the DOS of these much larger graphene planes with randomly distributed defects reveal the salient features that persist after taking into account the randomness character of the disorder.
In Fig.\ref{fig7}, the total DOS of large graphene planes containing $n_d=1$\% of SW, 585 and 555-777 divacancies, computed using the recursion method~\cite{Haydock}, is compared with the total DOS previously obtained for a $7\times7$ supercell containing one defect ($n_d\sim 1$\%). A first observation is that the DOS of random disordered systems is much smoother than the one corresponding to a supercell\cite{broadeningdos}. In the random case, most of the peaks have disappeared except the ones close to the Dirac point (In Fig.\ref{fig7}, the Dirac point has been set to zero.). The broadening due to the distribution disorder is more efficient in energy regions containing a lot of bands. Close to the Dirac point, the quantity of bands is less dense preserving the defect-induced resonances. Secondly, the position of resonance energy peaks corroborate the previous analysis carried out on supercell band structures. The presence of such resonant states induced by local defects is a well known issue~\cite{Wehling,PRLKubo2011,Mesaros,GTrambly,CockaynePRB2011}. The DOS of randomly disordered graphene confirm that the electron transport in an energy region around $E=0.35$ eV is expected to be degraded for SW defects, whereas hole transport should be altered around $E=-0.35$ eV for 585 divacancies, and finally that 555-777 divacancies exhibit several resonance energies around $E=0.6,-0.8,-2.1$ eV which should also lead to reduced transport performances.\\
The case of a mixture of defects is now explored by considering graphene planes containing, half SW half 585 divacancy (SW/585), then half SW half 555-777 divacancy (SW/555-777), and finally half 585 half 555-777 divacancies (585/555-777). The corresponding DOS of these systems for $n_d=1$\% of defects in total are plotted in Fig.\ref{fig8}. These DOS of graphene planes containing a mixture of two types of defects are compared also with the DOS of graphene planes containing a single type of defect separately (i.e. $n_d=0.5$\%). It is obvious that the features observed in the DOS of the mixed systems are roughly the sum of the individual features of each defect type. The particular case of SW/585 is interesting in the sense that the resonance peaks of these two defects are almost symmetric with respect to the Dirac point. Moreover, since the resonance peaks are relatively close to the initial Dirac point, they tend to overlap at this special point and yield to an increase of the DOS at the Dirac point. In this special situation, there is no more a clear minimum of DOS associated with the Dirac point. By adding other types of defects, a large increase of the DOS at the Dirac point can be foreseen. This is actually what is observed for highly defective or \textit{amorphous} graphene membranes~\cite{Lherbier-amorphous,Kapko,Holmstrom}. But this DOS increase comes from resonance states mainly localized around the defects which will therefore not participate to the transport of charge carriers but will rather degrade it.

\subsection{Kubo-Greenwood methodology}

The transport properties of large graphene planes containing structural defects are calculated using an efficient real-space order-N Kubo-Greenwood method~\cite{Roche97,Mayou95,Mayou2007,Triozon2002,Roche99,Ishii}. This very efficient technique gives a direct access to the main transport quantities in the semiclassical regime as well as in the quantum regime in which all multiple scattering events and interferences are retained. All quantities at energy $E$ are extracted or derived from the wave packet dynamics. The latter is characterized by the time-dependent diffusivity $D(E,t)=\Delta R^2(E,t) / t $ where $\Delta R^2 = \Delta X^2 + \Delta Y^2 $ is the mean quadratic spreading of wave packets. The mean quadratic spreading along a given direction is evaluated from the Hamiltonian and the corresponding position operator as $\Delta X^2 (E,t) = \rm{Tr}[\delta (E-\hat{H}) \vert \hat{X}(t) - \hat{X}(0)\vert^2] / \rm{Tr}[\delta (E-\hat{H})]$. $\rm{Tr}$ is the trace over ${\rm\textit{p}_z}$ orbitals and $\rm{Tr}[\delta (E-\hat{H})] / S = \rho(E)$ is the total DOS (per unit of surface). The two position operators $\hat{X}(t)$ and $\hat{Y}(t)$ are expressed in the Heisenberg representation ($\hat{X}(t) = \hat{U}^{\dagger}(t)\hat{X}(0)\hat{U}(t)$) and the time evolution operator $\hat{U}(t)=\Pi_{n=0}^{N-1}\exp(i\hat{H}\Delta t/\hbar)$, with $\Delta t$ the chosen time step, is computed with a Chebyshev polynomial expansion method~\cite{Roche97,Mayou95,Mayou2007,Triozon2002,Roche99,Ishii}. Calculations are performed for several initial random phase wave packets, and for a total elapsed time $t \approx 3.6$ps split in three parts with different time steps, $\Delta t_1 \approx 1.32$fs, $\Delta t_2 \approx 26.33$fs and $\Delta t_3 \approx 52.66$fs. The size of simulated graphene planes is $L_x\times L_y=300\times250\text{nm}^{2}\sim0.074 \mu \text{m}^{2}$ (\textit{i.e.\ }$2.8\times10^6$ atoms), large enough to avoid finite size effects while periodic boundary conditions are applied to ensure continuity of the propagation of the wave packets. In the Kubo-Greenwood formalism, the different transport regimes can be inferred from the time dependence of the diffusivity coefficient $D(E,t)$. The wave packet velocity $v(E)$ can be extracted for instance from the short time behavior of the diffusivity (\textit{ballistic regime}), $D(E,t) \sim v^2(E)t$, while the elastic mean free path $\ell_{e}(E)$ is estimated from the maximum of the diffusivity (\textit{diffusive regime}), $D_{\text{max}}(E)=2v(E)\ell_{e}(E)$. For two-dimensional systems, the Kubo-Greenwood conductivity equation is given by
\begin{equation}
\sigma(E_F,T_F)=-\int^{+\infty}_{-\infty}dE' \frac{\partial f(E',T_F)}{\partial E'} \sigma(E',0\text{K}) \label{eqsigmaKubo}
\end{equation}
where $E_F$ is the Fermi energy and $f(E,T_F)$ is the Fermi-Dirac distribution function with $T_F$ the associated Fermi-Dirac temperature. For $T_F=0$K, the Kubo-Greenwood conductivity is defined by 
\begin{equation}
\sigma(E_F,T_F=0\text{K})=\frac{1}{4}e^{2} \rho(E_F) \lim_{t\rightarrow \infty} \frac{\partial }{\partial t}\Delta R^2(E_F,t)\label{eqsigmaKuboT0K}
\end{equation}
\begin{figure}[b]
\begin{center}
\leavevmode
\includegraphics[width=1.0\columnwidth]{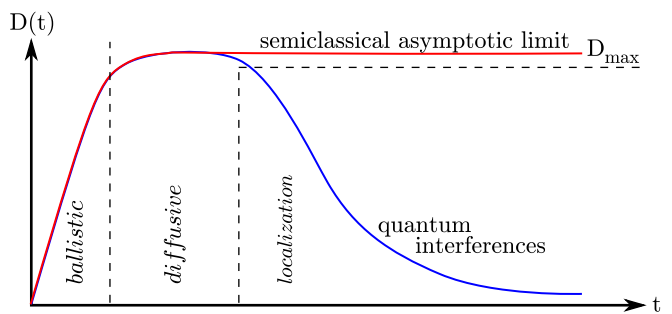}
\caption{(color online). Sketch of the typical behavior of $D(t)$ showing the three possible transport regime including the localization regime where the diffusivity decreases due to quantum interferences. In the semiclassical picture, $D(t)$ stays in the diffusive regime [i.e. $\lim_{t\mapsto\infty} D(t)=D_{\text{max}}$].}
\label{fig9}
\end{center}
\end{figure}
In these two last equations, the Fermi level ($E_F$) has to be thought as an energy that can be tuned with a virtual electrostatic gate, so that the transport is accessible in a large energy window around the equilibrium position of the Fermi level. To avoid any confusion however, we set $E_F\equiv E$ in the following and reserve the notation $E_F$ to the equilibrium Fermi energy  at zero gate voltage (when the system is charge neutral). Note that we ignore the effects of screening by charge accumulation, which have been shown to be negligible in section~\ref{sectionabinitio}.\\
In the semiclassical transport picture, the asymptotic behavior of $\Delta R^2(E,t)$ at long time is linearly proportional to $t$ (\textit{diffusive regime}). Therefore, the derivative with respect to the time in Eq.\ref{eqsigmaKuboT0K} can be replaced by a simple division by the time. The diffusivity coefficient is thus introduced as $D(t)=\Delta R^2(t)/t$ in the Kubo-Greenwood conductivity formula to obtain the expression of the semiclassical conductivity
\begin{eqnarray}
\sigma_{sc}(E,0\text{K})&=&\frac{1}{4}e^{2} \rho(E) \lim_{t\rightarrow \infty} D(E,t) \label{eqsigmaKuboT0Kbis}\\
\sigma_{sc}(E,0\text{K})&=&\frac{1}{4}e^{2} \rho(E) D_{\text{max}}(E)
\end{eqnarray}
Finally, following the scaling theory~\cite{LeeRamakrishnan,LeeFisher,EversMirlin}, the localization length $\xi(E)$ can be determined from the semiclassical transport length scales as,
\begin{equation}
\xi(E)=\sqrt{2}\ell_{e}(E)\exp\left(\frac{\pi h\sigma_{sc}(E,0\text{K})}{2e^{2}}\right)\label{ksiKubo}
\end{equation}
In the quantum regime, interferences in scattering paths survive and a decrease of $D(t)$ is therefore expected (\textit{localization regime}). A sketch in Fig.\ref{fig9} illustrates the typical behavior of the diffusivity coefficient as a function of time, outlining the various possible transport regimes.

\begin{figure*}[ht!]
\begin{center}
\leavevmode
\includegraphics[width=1.0\textwidth]{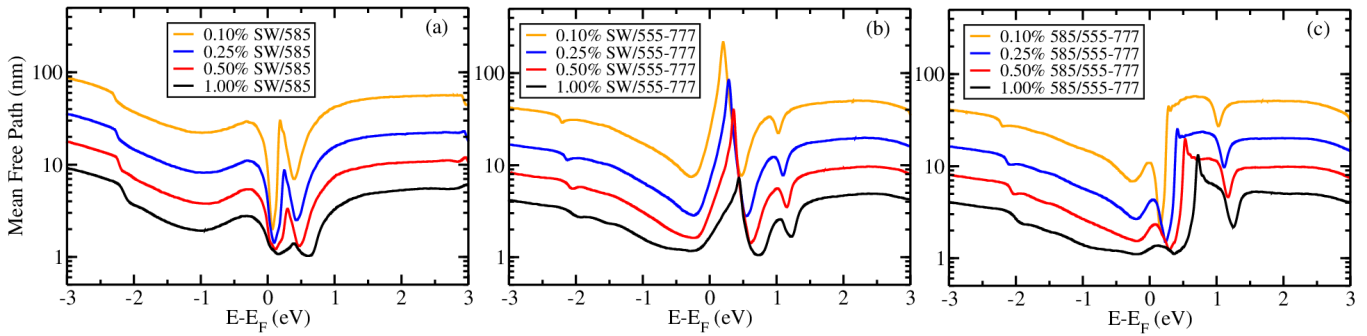}
\caption{(color online). Mean free paths in graphene planes with, (a) SW/585 defects, (b) SW/555-777 defects, and (c) 585/555-777 defects, for different defect concentrations ranging from $n_d=0.1$\% to $1.0$\%.}
\label{fig10}
\end{center}
\end{figure*}

\subsection{Semiclassical transport - Diffusive regime}\label{sec_semiclassical}

The semiclassical transport regime excludes all quantum interferences effects and corresponding localization phenomena (weak and strong localization regimes) and simply captures scattering events within the diffusive regime, as for example derived within the self consistent Born approximation~\cite{Shon,Ando}. Therefore, to extract the semiclassical transport length scales, the Kubo-Greenwood formalism is restricted to the diffusive regime limit, \textit{i.e.}\ the maximum of the diffusivity coefficient $D_{\text{max}}$ as shown in Fig.\ref{fig9} (red line). In the following, the semiclassical transport properties of graphene planes containing various types and densities of structural defects will be described. First, the elastic mean free paths will be examined, then the mobilities, and finally the semiclassical conductivities will be depicted.

\begin{center}
\textit{Mean Free Path}\\ 
\end{center} 

In Fig.\ref{fig10}, the calculated mean free paths is presented for graphene planes in which SW/585 defects, SW/555-777 defects, or 585/555-777 defects have been incorporated, with different defect concentrations ranging from $n_d=0.1$\% to $1.0$\%. The mean free paths exhibit strong variations in energy, with dips associated to the defect resonance energies (bumps in the DOS). Note that from now, Fermi energy is always set to zero. The largest and fastest variation is obtained in an energy window corresponding to the 585 defect resonance energy, which appears just at the right side of the Fermi energy in case of SW/585 and 585/555-777 systems. The maximum difference in mean free path is about one order of magnitude. A rapid variation of the transport length scales with energy (or gate voltage) offer interesting perspectives for engineering mobility gaps~\cite{Roche2011}. Actually, for the specific needs of logic devices which require an on and an off state, the case of 585/555-777 systems might be the most efficient one because the shape of the corresponding curve $\ell_{e}(E)$ is the closest to a stepwise function. For $n_d=0.1$\%, the maximum calculated mean free paths for all different defective systems are in the range of $[60,200]$nm whereas the minimal ones lie within $[2,10]$nm. For $n_d \geq 1$\%, the mean free paths are found to be smaller than $10$nm for any energy in the window considered here (\textit{i.e.}\ $[-3,+3]$eV). Finally, regarding the variation of the mean free path with defect densities, better ratios between minimum and maximum mean free paths are expected for defect concentrations $n_d<0.1$\%. Unfortunately, smaller defect concentrations are not accessible to our simulations due to computational limitations.

\begin{figure*}[h!]
\begin{center}
\leavevmode
\includegraphics[width=1.0\textwidth]{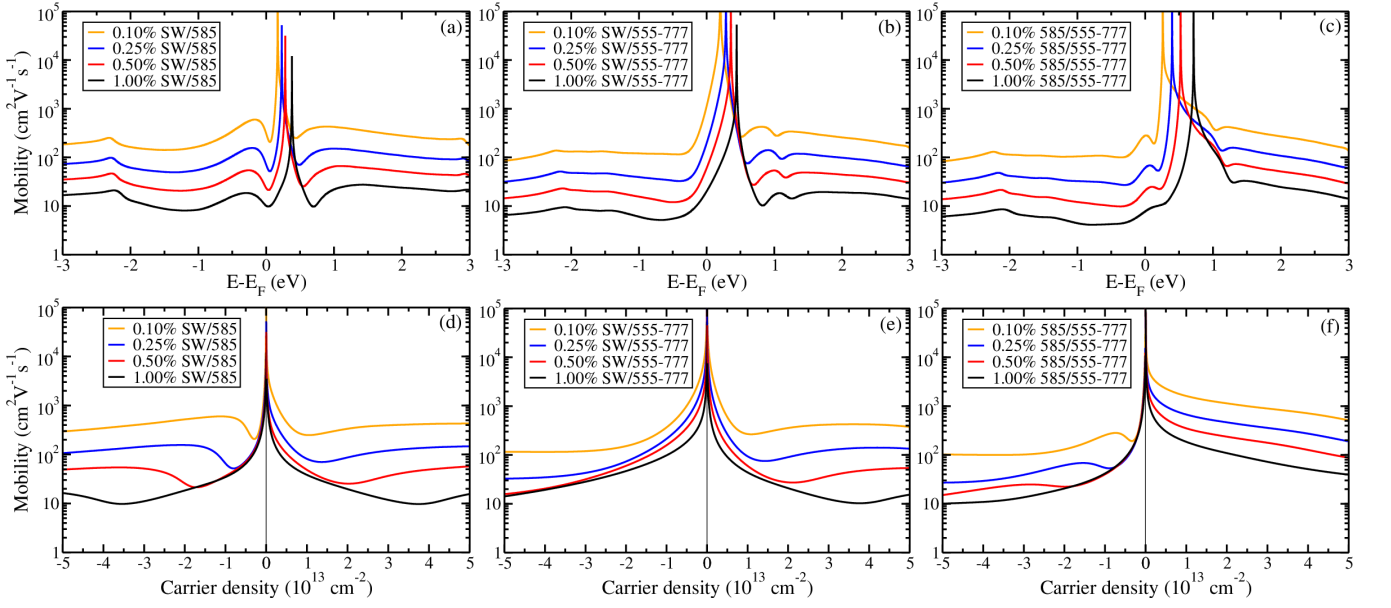}
\caption{(color online). Mobilities as a function of charge carrier energy (upper panels) and as a function of charge carrier density (lower panels) of graphene planes with, (a)(d) SW/585 defects, (b)(e) SW/555-777 defects, and (c)(f) 585/555-777 defects, for different defect concentrations ranging from $n_d=0.1$\% to $1.0$\%.}
\label{fig11}
\end{center}
\end{figure*}

\begin{figure*}[h!]
\begin{center}
\leavevmode
\includegraphics[width=1.0\textwidth]{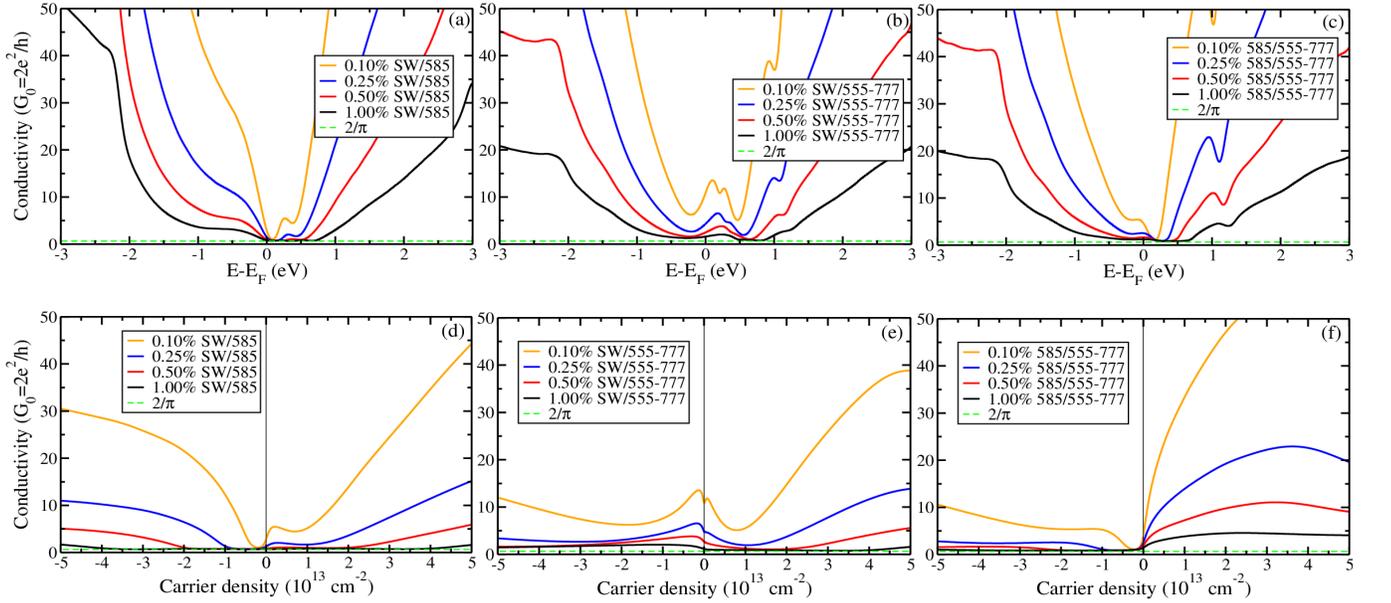}
\caption{(color online). Conductivities as a function of charge carrier energy (upper panels) and as a function of charge carrier density (lower panels) of graphene planes with, (a)(d) SW/585 defects, (b)(e) SW/555-777 defects, and (c)(f) 585/555-777 defects, for different defect concentrations ranging from $n_d=0.1$\% to $1.0$\%.}
\label{fig12}
\end{center}
\end{figure*}

\begin{center}
\textit{Mobility}\\ 
\end{center}

In Fig.\ref{fig11}, the mobilities as a function of carrier energy (upper panels) and of carrier density (lower panels) computed with a Fermi-Dirac temperature of $300$K are illustrated. The mobility ($\mu$) is calculated using the conventional definition $\mu(E,T_F)=\sigma_{sc}(E,T_F)/e\,n(E,T_F)$ where $e$ is the elementary charge and $n(E,T_F)$ the charge carrier density (electrons or holes) computed by integrating the DOS from the Dirac point to the energy $E$. Since $\mu$ is inversely proportional to the charge carrier density ($n$), there is an obvious mathematical divergence of $\mu$ when $n$ tends to zero (close to the Dirac point) as observed in Fig.\ref{fig11}. In each of the three cases (SW/585, SW/555-777, and 585/555-777), the p-doping character induced by divacancies is clearly illustrated by the continuous shift of the mobility peak with respect to the Fermi level as $n_d$ increases (upper panels). As for the mean free paths, some dips are also visible even if they seem not be as deeper because the logarithm scale used in Fig.\ref{fig11} run over five orders of magnitudes. The fastest variation is again obtained around the resonance energies associated to 585 defects and yields to a change of mobility values between two and three orders of magnitudes. For the largest defect concentration ($n_d=1$\%), the mobility can drop below $10$cm$^{2}$V$^{-1}$s$^{-1}$ but keeping a value of a few ten thousands of cm$^{2}$V$^{-1}$s$^{-1}$ at the Dirac point. In the lower panels of Fig.\ref{fig11}, the mobilities are plotted versus the electron and hole carrier densities measured with respect to the Dirac point. Therefore, the mobility peaks are all aligned. Variations with the carrier density are found to be smoother than with the carrier energy, which is due to the non linear dependence between $n$ and $E$ ($n(E)\propto E^2$). Nevertheless, the variation of hole mobility close to the Dirac point is reasonably sharp in case of 585/555-777 defective systems. Far away enough from the Dirac point, the mobility generally tends to a constant value meaning that the conductivity is linearly dependent to the charge carrier in these energy regions.

\begin{center}
\textit{Semiclassical Conductivity}\\ 
\end{center}

In Fig.\ref{fig12}, the semiclassical conductivities ($\sigma_{sc}$) are plotted as a function of carrier energy and density and for $T_F=300$K. As for mean free paths and mobilities, the conductivities exhibit dips at resonance energies. The value of conductivity is minimal at those energies, which indicates that a minimal conductivity can be obtained for an energy different to the Dirac point. There is also a minimum value of conductivity associated to the Dirac point which is visible for Fermi-Dirac temperature equal to $0$K (although not shown here). This minimum is very sharp and thus easily broadened by the Fermi-Dirac temperature. Nevertheless, for SW/555-777 systems, this minimum associated to the Dirac point is still observable in Fig.\ref{fig12}.b for $n_d=0.1$\% ($E\sim 0.2$eV).\\ When the defect concentration increases, the conductivity at the defect resonances reaches the semiclassical limit of minimum conductivity $\sigma_{sc}^{min}=2\,G_0/\pi=4e^2/\pi\,h$, leading to the progressive formation of a plateau of conductivity (see ref~\cite{Cresti} for a discussion on the minimum conductivity). Consequently, the conductivity curves present a lot of different behavior versus the energy. These different behaviors (constant, linear, quadratic,...) are even more obvious in the lower panels of Fig.\ref{fig12}, where conductivities are plotted versus the carrier density. Indeed, for $n_d=0.1$\%, depending on the defective system and on the type of carriers, either a linear behavior can be observed (for electrons of SW/585 and in a smaller extend for electrons of SW/555-777), or a quadratic behavior (for electrons and holes close to the Dirac point and more precisely around the defect resonances of all three type of defective systems), or again a sub-linear behavior (for holes of SW/585 and electrons of 585/555-777). For $n_d=1.0$\% of defects, the conductivity even displays an almost constant behavior over a large range of carrier density since the conductivity has reached the semiclassical conductivity limit $\sigma_{sc}^{min}$. The fact that all these different behaviors for $\sigma_{sc}(n)$ are present in these defective graphene systems is an interesting observation for the current puzzling interpretation of $\sigma_{sc}(n)$~\cite{Stauber,Yan2011}.\\
The semiclassical electronic transport in graphene containing various mixtures of structural defects has been investigated in this paragraph. The structural defects, which break locally the symmetry of the honeycomb lattice, have been found to yield to resonance scattering at specific energies. These resonant defect levels induce peaks in the DOS which increase with the defect density. These resonant states conduct systematically to a degradation of all transport properties examined (mean free paths, mobilities, and conductivities). This degradation occurring at specific energies, constitute a fingerprint related to each type of defect separately even when several types of defects coexist in the system (provided that resonant levels are not to close in energy to overlap, which could be the case with other defects not studied here). The 585 divacancy produces a rather sharp resonance compared to the two other defects. In particular, the case of 585/555-777 defective system presents an interesting profile regarding variations of transport properties with respect to carrier energy and carrier density. Finally, large plateaus of minimum conductivity are predicted for sufficiently high concentration of defect ($n_d > 0.5$\%).\\
In the next section, the quantum regime of transport in such defective graphene systems will be discussed. In this transport regime, the quantum interferences in electron trajectory will be taken into account which yields to localization phenomena and thus an additional degradation of transport properties such as the conductivity.

\subsection{Quantum transport - Localization regime}

In the localization transport regime, quantum interferences can yield to an enhanced probability to be back-scattered. In a common picture, this effect can be viewed as constructive interferences between forward and backward electron scattering trajectories that form loops. This is possible only if the electron keeps its phase and thus, any source of phase decoherence such as electron-phonon scattering will destroy these interferences. This is the reason why such localization effects are usually observable at low temperature only. However, in graphene the electron-phonon scattering have been found to be extremely low~\cite{JHChen2008,Morozov}, allowing the localization phenomena to occur even at room temperature. Concerning the transport quantities, the quantum interferences induce a correction which can be described theoretically trough a Cooperon term~\cite{Montambaux_Akkermans}. For a two-dimensional system, this Cooperon correction yields for instance to a decrease of conductivity which scale logarithmically with the sample size (weak localization regime). For even stronger disordered systems, the scaling of the conductivity exhibits an exponential decay (strong localization ergime), and the electronic states become strongly localized such that the system behaves as an insulator from a transport properties point of view. Several theoretical and experimental studies on graphene have reported such a (semi)metal-insulator transition through a Anderson localization of electronic states which was cause either by hydrogen atoms~\cite{Bang10,Skrypnyk,Schubert,Grassi,Elias,Bostwick}, epoxy oxygen atoms~\cite{LeconteACS,Moser10,LecontePRB}, vacancies and other structural defects~\cite{PRLKubo2011,GTrambly,Uppstu,Haskins,Wu}, or even noble metal clusters~\cite{WeiLi11}.

\begin{figure*}[ht!]
\begin{center}
\leavevmode
\includegraphics[width=1.0\textwidth]{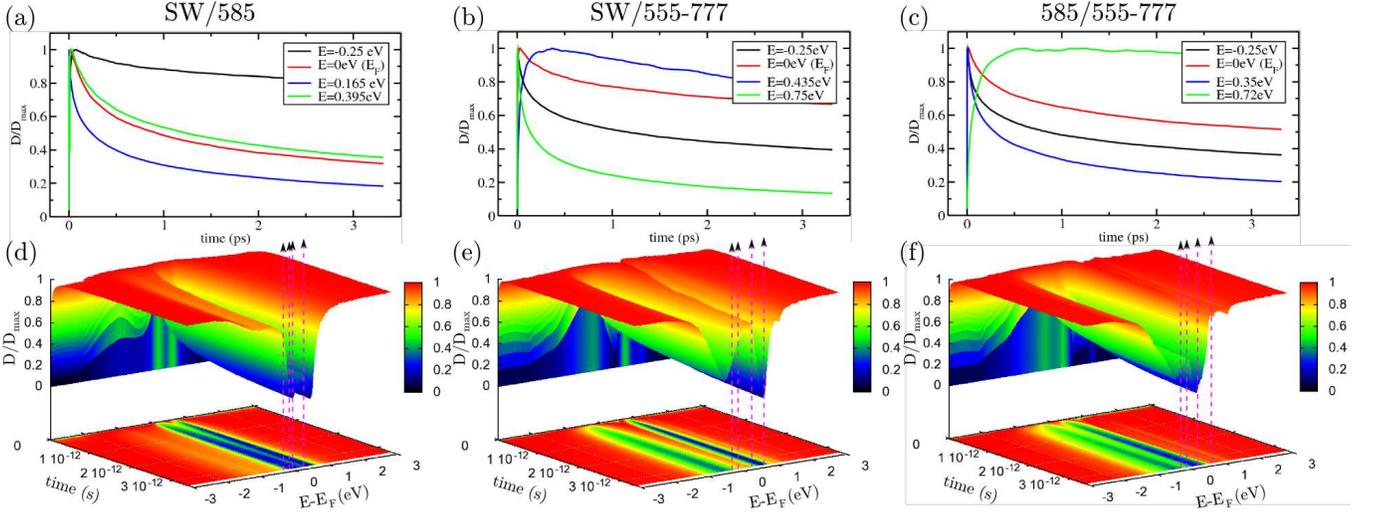}
\caption{(color online). Normalized diffusivity as a function of time for four selected energies (upper panels), and $3$D plots as a function of time and energy (lower panels), of graphene planes with, (a)(d) SW/585 defects, (b)(e) SW/555-777 defects, and (c)(f) 585/555-777 defects, for a defect concentration $n_d=1.0$\%.}
\label{fig13}
\end{center}
\end{figure*}

\begin{center}
\textit{Diffusivity}\\ 
\end{center}

In our real-space Kubo-Greenwood calculations, this degradation of transport quantities caused by the quantum interferences can be directly observed through the time dependence of the diffusivity coefficient. The Figs.\ref{fig13}.a-c illustrate this time dependence of normalized diffusion coefficients ($D(t)/D_{\text{max}}$) for four selected energies (including the Fermi energy), while Figs.\ref{fig13}.d-f present a $3D$ plot of the same quantity for a large energy window of the spectrum. The four energies selected in the Figs.\ref{fig13}.a-c are pointed with vertical arrows in Figs.\ref{fig13}.d-f. The decreasing of $D(t)$ clearly reveals the emergence of localization effects. This decreasing is not homogeneous in the whole spectrum but stronger at the defect resonance energies. Actually, far away from these resonance energies, there is no more decreasing and the behavior of $D(t)$ at long time is a constant which is typical of a semiclassical diffusive regime.

\begin{center}
\textit{Localization Length}\\ 
\end{center}

\begin{figure*}[ht!]
\begin{center}
\leavevmode
\includegraphics[width=1.0\textwidth]{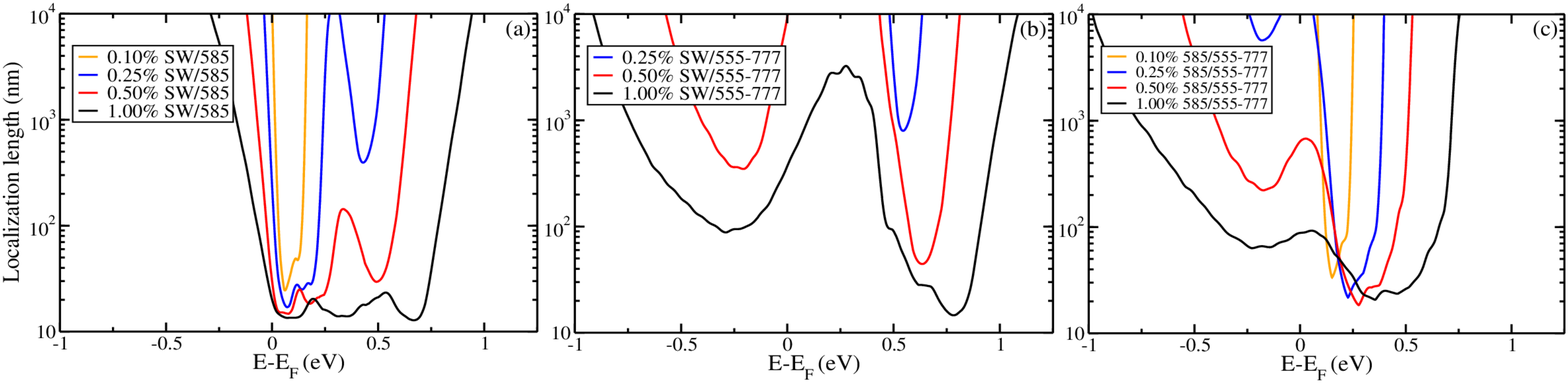}
\caption{(color online). Localization lengths, evaluated using Eq.\ref{ksiKubo}, in graphene planes with, (a) SW/585 defects, (b) SW/555-777 defects, and (c) 585/555-777 defects, for different defect concentrations ranging from $n_d=0.1$\% to $1.0$\%.}
\label{fig14}
\end{center}
\end{figure*}

Since localization phenomena at some specific energies have been observed, the corresponding localization length ($\xi$) can be estimated using Eq.\ref{ksiKubo} (see Fig.\ref{fig14}). For the highest defect concentration ($1.0$\%), the smallest localization lengths, obtained exactly at the defect resonant energies, are $\xi=10$-$20$nm about. At these energies, the conductivity reaches the semiclassical lower limit (\textit{i.e.}\ $\sigma_{sc}^{min}$), and therefore, Eq.\ref{ksiKubo} becomes $\xi(E)=\sqrt{2}\ell_{e}(E)\exp(2)= 10.5\ell_{e}(E)$ with the smallest value obtained for $\ell_{e}$ being around $1$nm. Away from the resonance energies, the localization length increases very rapidly mostly driven by the conductivity which appears in the exponential function. This exponential variation with the conductivity is a common feature of two-dimensional systems which make the experimental observation of localization generally difficult even if the mean free path is very small.\\
\\

\begin{center}
\textit{Extension of the real-space Kubo-Greenwood methodology}\\ 
\end{center}

In the section~\ref{sec_semiclassical}, the Kubo-Greenwood methodology has been used to describe semiclassical transport properties using quantities extracted only from the diffusive regime (maximum of $D(t)$). In the previous paragraph, even the localization lengths have been extracted from semiclassical quantities using the scaling theory of localization (Eq.\ref{ksiKubo}). The reason why the Kubo-Greenwood formalism is usually restricted to the diffusive regime, is because this is a \textit{bulk} formalism in the sense that the framework is devoted to the description of the intrinsic properties of a material. For instance, within the Kubo-Greenwood formalism, the calculated semiclassical conductivity is an intensive quantity contrarily to the conductance which is extensive and depends on the system geometry. Hence, the Kubo-Greenwood conductivity formula is usually employed in its thermodynamic limit $t\rightarrow \infty$ (Eqs.\ref{eqsigmaKuboT0K}-\ref{eqsigmaKuboT0Kbis}). This infinite time limit imposes to restrict the analysis of $D(t)$ to its asymptotic value which can be either a constant (purely diffusive regime, $\sigma=\sigma_{sc}$) or zero (completely localized regime, $\sigma=0$). On the contrary, the Landauer-B\"{u}ttiker transport approach, is better suited to describe extensive transport properties such as the conductance ($G(L)$). In this last formalism, the presence of leads introduces a finite length $L$ of the system which allows scaling analysis describing therefore all intermediate regimes. Even if there is \textit{a priori} no clear definition of length in the Kubo-Greenwood approach since the system is periodic (bulk), the present real-space implementation of this transport formalism provides a complete description of the dynamics of propagation, going from the ballistic regime at very short time simulations to quasi-diffusive and diffusive regimes as well as the weak and strong localization regimes at longer time. In fact, in these real-space simulations, the computed spreading of wave packets can be seen as an average time-dependent distance probed by the propagating charge carriers. To allow a scaling analysis within the real-space Kubo-Greenwood approach, the introduction of a characteristic length scale $L(t)$ defined as the diameter of a fictitious circle formed by the spreading of wave packets is therefore proposed. 
\begin{equation}
L(E,t) = 2 \sqrt{\Delta R^2(E,t)} \label{eqlengthkubo}
\end{equation}
This new approach, which will give access to length dependent transport quantities and allow to describe more in details the localization phenomena, will be verified and justified in the following by comparing the results with the predictions of the scaling theory of localization. Before that, the definition of the diffusion coefficient $D(t)$ has to be reexamined.
\\

\begin{figure}[ht!]
\begin{center}
\leavevmode
\includegraphics[width=1.0\columnwidth]{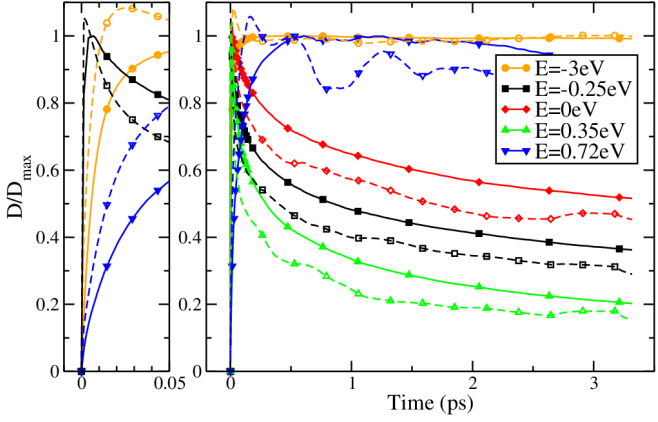}
\caption{(color online). Normalized diffusivity $D(t)$ in graphene planes with $1.0$\% of 585/555-777 defects, as a function of time for five selected energies. Solid lines are obtained with $\Delta R^2(t)/t$ whereas dashed lines are obtained with $\frac{\partial }{\partial t}\Delta R^2(t)$. Both expressions are divided by $D_{\text{max}}$ found for $\Delta R^2(t)/t$.}
\label{fig15}
\end{center}
\end{figure}

As pointed out by Eq.\ref{eqsigmaKuboT0K}, the general definition of the diffusion coefficient is given by $D(t)=\frac{\partial }{\partial t}\Delta R^2(t)$. In the diffusive regime, the quadratic spreading of wave packets being linearly proportional to the time ($\Delta R^2(t)\propto t$), its first derivative reduces exactly to the constant value $D(t)=\Delta R^2(t)/t$. For numerical convenience, it is more efficient to use the definition $D(t)=\Delta R^2(t)/t$. Indeed, the first derivative is more computationally demanding and its numerical evaluation can produce large fluctuations from which it is difficult to extract semiclassical transport properties. In the localization regime, since $D(t)$ is no more a constant, the general definition $D(t)=\frac{\partial }{\partial t}\Delta R^2(t)$ should be used. However, although using $D(t)=\Delta R^2(t)/t$ is in principle incorrect in the localization regime, it is nevertheless interesting to compare the two definitions. In Fig.\ref{fig15}, the diffusivity coefficient is plotted using either $D(t)=\Delta R^2(t)/t$ (solid lines) or $D(t)=\frac{\partial }{\partial t}\Delta R^2(t)$ (dashed lines). For sake of comparison, both definitions are normalized with the same value of maximum of diffusivity ($D_{\text{max}}$) found for $D(t)=\Delta R^2(t)/t$. For an energy far from the resonance energies, \textit{e.g.}\ $E=-3eV$ for which no localization effects are observed, the two curves (solid and dashed lines) are almost superimposed. Both evaluations of $D(t)$ at long time, deep in the diffusive regime, gives thus the same constant value. The only difference is a bump obtained for the diffusivity evaluated with the numerical derivative. This bump appears at the transition between the ballistic and the diffusive regime. The bump is followed by a small decrease of $D(t)$ before the saturation is obtained. In principle, there is no reason to observe a decreasing of $D(t)$ in a purely diffusive regime as shown in Fig.\ref{fig9}. This bump is thus just an artifact produced by small numerical errors in $\Delta R^2(t)$ then enhanced by its first derivative calculation~\cite{note}. In the localization regime, which is clearly predicted for energies $E=-0.25, 0, 0.35$eV, Fig.\ref{fig15} shows that the diffusivity computed with $D(t)=\Delta R^2(t)/t$ overestimates the one derived with $D(t)=\frac{\partial }{\partial t}\Delta R^2(t)$. Nevertheless, the overestimation in the localization regime being almost constant with increasing time, the general behavior is still correct when $D(t)$ is evaluated with $D(t)=\Delta R^2(t)/t$. Moreover, the curves obtained with that definition are much smoother. The definition $D(t)=\Delta R^2(t)/t$ could therefore be conserved even in the localization regime, keeping in mind that it slightly overestimates the true value. With such a clarification for the definition of diffusivity in the localization regime, the scaling analysis of the conductivity in the quantum regime can be performed.

By using Eq.\ref{eqlengthkubo}, transport quantities can be expressed as a function of length ($L$) instead of as a function of time ($t$). To check the validity of this presumed definition of $L$, the scaling of the conductivity can be compared to the predictions of the scaling theory. In particular, this theory predicts the length dependence of quantum corrections (Cooperon term) to the conductivity due to localization effects. For two-dimensional systems, these corrections are logarithmic~\cite{Montambaux_Akkermans} and read as,
\begin{eqnarray}
\sigma(E,L) &=& \sigma_{sc}(E) - \frac{2e^2}{h\pi} \ln \left( \frac{L}{\sqrt{2}\ell_{e}(E)} \right) \nonumber\\
            &=& \sigma_{sc}(E) - \frac{G_0}{\pi} \ln \left( \frac{L}{\sqrt{2}\ell_{e}(E)} \right) \label{eqscalinglaw}
\end{eqnarray} 

\begin{figure}[ht!]
\begin{center}
\leavevmode
\includegraphics[width=1.0\columnwidth]{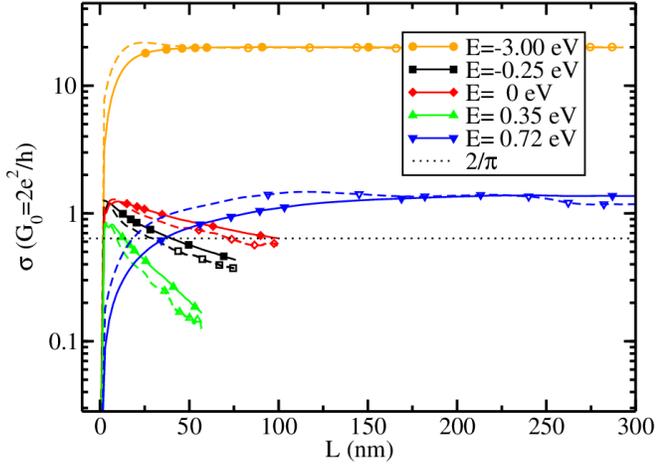}
\caption{(color online). Quantum conductivity $\sigma(E,L)$ in graphene planes with $1.0$\% of 585/555-777 defects, as a function of length for five selected energies. We use $D(t)=\Delta R^2(t)/t$ (solid lines) and $D(t)=\frac{\partial }{\partial t}\Delta R^2(t)$ (dashed lines) for computing $\sigma(L)$. The horizontal dotted line gives the semiclassical limit $\sigma_{sc}^{min}=2/\pi\,G_0=4e^2/\pi\,h$.}
\label{fig16}
\end{center}
\end{figure}

In Fig.\ref{fig16}, using Eq.\ref{eqlengthkubo}, the quantum conductivity as a function of length ($\sigma(L)$) is presented employing either $D(t)=\Delta R^2(t)/t$ (solid lines) or $D(t)=\frac{\partial }{\partial t}\Delta R^2(t)$ (dashed lines). The horizontal dotted line denotes $\sigma_{sc}^{min}$. As for the diffusivity (Fig.\ref{fig15}) both definitions of $D(t)$ lead to similar conductivity curves but the use of $D(t)=\Delta R^2(t)/t$ produces smoother curves and a slight overestimation in the localization regime. Localization effects are absent for $E=-3$eV as evidenced by the saturation of $\sigma(L)$ to its semiclassical asymptotic constant value $\sigma_{sc}$. For other energies (especially for $E=-0.25, 0, 0.35$eV), localization effects induce a decreasing of quantum conductivity with increasing length. The semiclassical value $\sigma_{sc}$ (which is maximum of $\sigma(L)\equiv\sigma_{\text{max}}$) is always found to be greater than $\sigma_{sc}^{min}=2G_0/\pi$, while quantum conductivity can decrease below such a value for sufficiently long lengths and for energies close enough to the defect resonance energies. Finally, Fig.\ref{fig16} exhibits a maximum computed length which is different for each energy. This is explained by the length scale (Eq.\ref{eqlengthkubo}) which is energy dependent, thus for a fixed total elapsed time, the length probed by the spreading of wave packets can be different.

\begin{figure}[ht!]
\begin{center}
\leavevmode
\includegraphics[width=1.0\columnwidth]{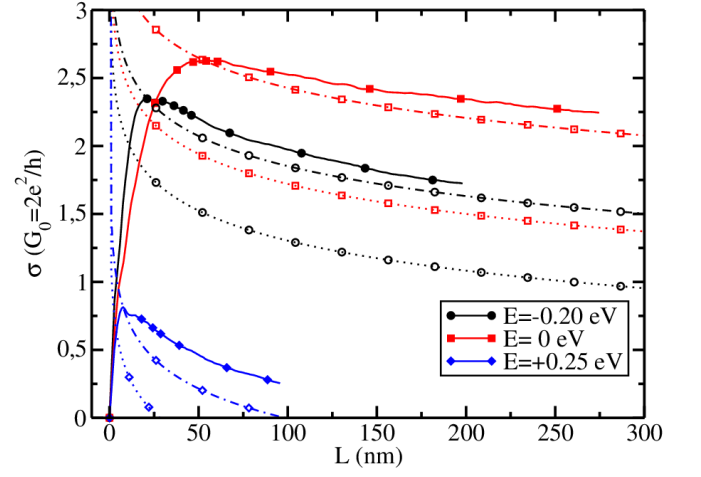}
\caption{(color online). Quantum conductivity in graphene planes with $0.25$\% of 585/555-777 defects, as a function of length and for three selected energies. Solid lines are obtained using $D(t)=\Delta R^2(t)/t$. The dotted lines are obtained using Eq.\ref{eqscalinglaw} the and dashed-dotted lines using Eq.\ref{eqscalinglaw2} in which the new length scale $l_{\text{diff}}$ has been introduced.}
\label{fig17}
\end{center}
\end{figure}

In Fig.\ref{fig17}, the validity of the definition of the length $L$ (Eq.\ref{eqlengthkubo}) is checked by scrutinizing if the quantum conductivity decay (solid lines) follows a logarithmic behavior as predicted by the scaling theory (Eq.\ref{eqscalinglaw}). The dotted lines correspond to Eq.\ref{eqscalinglaw}. The dashed-dotted lines correspond to a modified version of Eq.\ref{eqscalinglaw} in which a new length scale ($l_{\text{diff}}$) has been introduced replacing the mean free path, 
\begin{equation}
\sigma(E,L) = \sigma_{sc}(E) - \frac{G_0}{\pi} \ln \left( \frac{L}{l_{\text{diff}}(E)} \right) \label{eqscalinglaw2}
\end{equation} 
with $l_{\text{diff}}$ defined as the length corresponding to $\sigma_{\text{max}}$.\\
In the review of Lee and Ramakrishnan~\cite{LeeRamakrishnan}, the equation of weak localization correction in the scaling theory is derived using the mean free path as the characteristic length scale of the diffusive regime. According to this definition, when the length $L$ is equal to the mean free path $\ell_{e}$, the conductivity is  maximal and corresponds to the semiclassical conductivity ($\sigma_{sc}\equiv\sigma_{\text{max}}$), and for $L>\ell_{e}$ weak localization develops. In contrast, using Eq.\ref{eqlengthkubo}, the obtained semiclassical conductivity ( or $\sigma_{\text{max}}$) occurs at a length scale different from $\ell_{e}$. For instance in Fig.\ref{fig17}, for an energy $E=0$eV ($E=E_F$), the maximum of conductivity occurs at $L=l_{\text{diff}}\sim53$nm, which is about ten times greater than the mean free path computed at this energy ($\ell_{e}\sim4$nm). Therefore, the definition of length Eq.\ref{eqlengthkubo} might be questioned and argued to be misleading. But, considering the scaling in time for which there is no ambiguity, there is also a factor of ten between the calculated mean scattering time $\tau$ and the time $t$ for which the diffusivity coefficient reaches its maximum value. Actually, in a semiclassical picture, this is consistent with the fact that to reach completely the diffusive regime (saturation of $D(t)$), the charge carriers needs to encounter several scattering events which corresponds thus to a total traveling distance of a few times the mean free path. The same idea also explains that the charge carriers need several scattering events before the weak localization corrections start to present a clear effect. Indeed, these corrections are due to constructive interferences between forward and backward scattering paths which form loop trajectories. The shortest loop trajectory is at least two times the mean free path in a semiclassical picture. The introduction of an additional length scale $l_{\text{diff}}$ is therefore meaningful for the interpretation of such quantum corrections. In Fig.\ref{fig17}, it is obvious that the conductivity decay follows the logarithmic behavior predicted by Eq.\ref{eqscalinglaw}. However, using the mean free path as a characteristic length scale in the formula does not allow to fit $\sigma(L)$. On the contrary, Eq.\ref{eqscalinglaw2} allows to more correctly predict the scaling of the conductivity and therefore to extrapolate to lengths $L$ longer than the ones computed. As a consequence, the localization length $\xi$ can be extracted with a better precision by replacing the mean free path $\ell_{e}$ by the diffusion length $l_{\text{diff}}$. Indeed, $\xi$ can be evaluated as the length $L$ for which the conductivity is equal to zero or equivalently for which the weak localization corrections are equal to the semiclassical conductivity.

\begin{eqnarray}
\sigma(E,L=\xi) &=& 0 = \sigma_{sc}(E) - \frac{G_0}{\pi} \ln \left( \frac{L}{l_{\text{diff}}(E)} \right) \nonumber \\
\xi(E) &=& l_{\text{diff}}(E)\exp\left(\frac{\pi \sigma_{sc}(E)}{G_0}\right)\label{ksiKubo2}
\end{eqnarray}

Compared to Eq.\ref{ksiKubo}, the localization lengths calculated using Eq.\ref{ksiKubo2} are roughly ten times greater for weak localization regime and about $4$-$5$ times greater for stronger localization regime. For instance, in Fig.\ref{fig17} for an energy $E=0.25$eV which corresponds exactly to the resonance of 585 divacancy and for which the localization effects are stronger, $\xi\sim25$nm obtained with Eq.\ref{ksiKubo} and $\xi\sim100$nm estimated with Eq.\ref{ksiKubo2}. Note however that in case of strong localization effects such as $E=0.25$eV in Fig.\ref{fig17} and for higher defect densities such as $1$\% (not shown here), there is a deviance between the logarithmic decay predicted by scaling theory for the weak localization corrections and the actual computed quantum conductivities. As already reported recently in~\cite{LecontePRB}, in such strong localization case, the weak localization corrections are no more suitable and an exponential decay has to be used to describe the scaling of $\sigma(L)$ which results in long tail corresponding to evanescent mode. Finally, in very strong localization regime, the localization length could alternatively be evaluated from the long time saturation of the quadratic spreading~\cite{Triozon2000} as
\begin{equation}
\xi(E) = \lim_{t\mapsto\infty} 2\sqrt{\Delta R^2(E,t)} = \lim_{t\mapsto\infty} L(E,t)
\end{equation}
Unfortunately, in our simulations, a complete saturation of the quadratic spreading has never been observed for the total elapsed time computed. 

\section{Conclusion}\label{sec3}

In conclusion, the electronic and transport properties of graphene planes containing structural defects such as, SW, 585 and 555-777 divacancies, has been investigated theoretically. First, \textit{ab initio} techniques have been used to study the geometry and the electronic band structures of single defects in a supercell. From these results, relaxed atomic positions and electronic defect potentials have been employed to build accurate TB models for each individual structural defects. Comparisons between \textit{ab initio} and TB band structures and DOS have shown a good agreement. The elaborated TB models have then been employed to compute DOS of large graphene planes containing randomly distributed structural defects of various natures and concentrations. The comparison of these DOS with the DOS obtained for supercell have put in evidence the smoothening effect of the randomness character in distribution and orientation of defect at a mesoscopic scale. Then, the computed DOS of graphene planes including a mixture of defects have illustrated that defect-induced resonant peaks can be assigned to individual defect separately demonstrating that those resonances constitute a fingerprint of each structural defect. Afterwards, using a Kubo-Greenwood method, the transport properties have been examined. The computed semiclassical transport quantities, such as mean free path, mobility, and conductivity have exhibited systematic degradations around resonant defect energies. In particular, it has been predicted that a minimum conductivity ($\sigma_{sc}^{min}=4e^2/\pi\,h$) can be obtained at an energy different than Dirac point, and that plateau of minimum conductivity are formed when increasing the defect density. Finally, a detailed analysis of the diffusivity deep into the localization regime has allowed to extend the use of the Kubo-Greenwood method to predict length dependence of the quantum conductivity ($\sigma(L)$). The latter exhibits a logarithmic decay with increasing lengths around defect resonance energies due to localization phenomena as expected by the scaling theory. An Anderson insulating behaviour is thus anticipated for graphene with $1\%$ of structural defect and associated localization lengths are expected to be in the order of a few tens nanometers. However, localization effects are not observed in all the spectrum but only around defect resonant energies complicating therefore its experimental observation. Transport measurements would thus require an efficient electrostatic gate to probe larger part of the spectrum and to align Fermi energy with these defect resonant energies. Additionally, transport characterization with different sizes of graphene samples exposed to the same amount of structural defects could be conducted in order to perform scaling analysis of the conductivity. Alternatively, magneto-transport experiments on a fixed size sample could also reveal localization phenomena induced by structural defects as well as low temperature transport measurements which should evidence a variable range hopping behaviour.

J.-C.C. and A.L. acknowledge financial support from the F.R.S.-FNRS of Belgium. S.M.-M.D. acknowledges funding from EPSRC (Grant n.$^{\circ}$ EP/G055904/1), X.D. acknowledges financial support from the FRIA. Parts of this work are connected to the Belgian Program on Interuniversity Attraction Poles (PAI6), to the ARC Graphene Nano-electromechanics (n.$^{\circ}$ 11/16-037) sponsored by the Communaut\'e Fran\c{c}aise de Belgique, to the European Union through the ETSF e-I3 project (grant n.$^{\circ}$ 211956), and to the NANOSIM-GRAPHENE Project (n.$^{\circ}$ ANR-09-NANO-016-01). Computational resources have been provided by the CISM of the Universit\'e catholique de Louvain (UCL).


\begin{thebibliography}{50}

\bibitem{NovoselovRMP}
K.S Novoselov, Rev. Mod. Phys. {\bf 83}, 837 (2011).

\bibitem{Zhang}
Y. Zhang, Y.-W. Tan, H.L. Stormer, and P. Kim, Nature {\bf 438}, 201 (2005).

\bibitem{FOrtman}
F. Ortman, A. Cresti, G. Montambaux, and S. Roche, Eur. Phys. Lett. {\bf 94}, 47006 (2011).

\bibitem{Bolotin}
K. Bolotin, K. Sikes, Z. Jiang, M. Klima, G. Fudenberg, J. Hone, P. Kim, and H.L. Stormer, Solid State Comm. {\bf 146}, 351 (2008).

\bibitem{Miao}
F. Miao, S. Wijeratne, Y. Zhang, U.C. Coskun, W. Bao, and C.N. Lau, Sience {\bf 317}, 1530 (2007).

\bibitem{Krasheninnikov}
A.V. Krasheninnikov, and F. Banhart, Nature Materials {\bf 6}, 723 (2007).

\bibitem{Charlier}
J.-C. Charlier, X. Blase, and S. Roche, Rev. Mod. Phys. {\bf 79}, 677 (2007).

\bibitem{Suenaga}
K. Suenaga, H. Wakabayashi, M. Koshino, Y. Sato, K. Urita, and S. Iijima, Nature Nanotechnology {\bf 2}, 358 (2007).

\bibitem{Lusk}
M.T. Lusk, D.T. Wu, and L.D. Carr, Phys. Rev. B {\bf 81}, 155444 (2010).

\bibitem{Cresti}
A. Cresti, N. Nemec, B. Biel, G. Niebler, F. Triozon, G. Cuniberti, and S. Roche, Nano Res. {\bf 1}, 361 (2008).

\bibitem{lherbier_dopant}
A. Lherbier, X. Blase, Y.M. Niquet, F. Triozon, and S. Roche, Phys. Rev. Lett. {\bf 101}, 036808 (2008).

\bibitem{Wang}
B. Wang, and S.T. Pantelides, Phys. Rev. B {\bf 83}, 245403 (2011).

\bibitem{Biel1}
C. G\'{o}mez-Navarro, P.J. De Pablo, J. G\'{o}mez-Herrero, B. Biel, F.J. Garcia-Vidal, A. Rubio, and F. Flores, Nature Materials {\bf 4}, 534 (2005). 

\bibitem{Biel2}
B. Biel, F.J. Garcia-Vidal, A. Rubio, and F. Flores, Phys. Rev. Lett. {\bf 95}, 266801 (2005).

\bibitem{JHChen2009}
J.-H. Chen, W.G. Cullen, C. Jang, M.S. Fuhrer, and E.D. Williams, Phys. Rev. Lett. {\bf 102}, 236805 (2009).

\bibitem{Libisch}
F. Libisch, S. Rotter, and J. Burgd\"{o}rfer, arXiv:1102.3848v1 (2011).

\bibitem{Lherbier08}
A. Lherbier, B. Biel, Y.M. Niquet, and S. Roche, Phys. Rev. Lett. {\bf 100}, 036803 (2008).

\bibitem{Banhart}
F. Banhart, J. Kotakoski, and A.V. Krasheninnikov, ACS NANO {\bf 5}, 26 (2011).

\bibitem{Kotakoski}
J. Kotakoski, A.V. Krasheninnikov, U. Kaiser, and J.C. Meyer, Phys. Rev. Lett. {\bf 106}, 105505 (2011).

\bibitem{CockaynePRB2011}
E. Cockayne, G.M. Rutter, N.P. Guisinger, J.N. Crain, P.N. First, and J. Stroscio, Phys. Rev. B {\bf 83}, 195425 (2011).

\bibitem{Cockayne}
E. Cockayne, arXiv:1106.6273v2 (2011).

\bibitem{Stone-Wales-paper}
A.J. Stone, and D.J. Wales, Chem. Phys. Lett. {\bf 128}, 501 (1986).

\bibitem{Ma}
J. Ma, D. Alf\`{e}, A. Michaelides, and E. Wang, Phys. Rev. B {\bf 80}, 033407 (2009).

\bibitem{Huertas-Hernando}
D. Huertas-Hernando, F. Guinea, and A. Brataas, Phys. Rev. B {\bf 74}, 155426 (2006).

\bibitem{GundoLee}
G.-D. Lee, C.Z. Wang, E. Yoon, N.-M. Hwang, D.-Y. Kim, and K.M. Ho, Phys. Rev. Lett. {\bf 95}, 205501 (2005).

\bibitem{Kim}
Y. Kim, J. Ihm, E. Yoon, and G.-D. Lee, Phys. Rev. B {\bf 84}, 075445 (2011).

\bibitem{Meyer}
J.C. Meyer, C. Kisielowski, R. Erni, M.D. Rossell, M.F. Crommie, and A. Zettl, Nano Lett. {\bf 8}, 3582 (2008).

\bibitem{Ugeda}
M.M. Ugeda, I. Brihuega, F. Hiebel, P. Mallet, J.-Y. Veuillen, J.M. G\'{o}mez-R\'{o}driguez, and F. Yndur\'{a}in, Phys. Rev. B {\bf 85}, 121402(R) (2012).

\bibitem{SIESTA}
J.M. Soler, E. Artacho, J.D. Gale, A. Garcia, J. Junquera, P. Ordejon, D. Sanchez-Portal, Journal of  Physics: Condensed Matter {\bf 14}, 2745 (2002).

\bibitem{Ceperley-Alder}
D.M. Ceperley, and B.J. Alder, Phys. Rev. Lett. {\bf 45}, 566 (1980).

\bibitem{Perdew-Zunger}
J.P. Perdew, and A. Zunger, Phys. Rev. B {\bf 23}, 5048 (1981).

\bibitem{Troullier-Martins}
N. Troullier, and J.-L. Martins, Phys. Rev. B {\bf 43}, 1993 (1991).

\bibitem{Artacho}
E. Artacho, D. S\'{a}nchez-Portal, P. Ordej\'{o}n, A. Garc\'{\i}a, and J.M. Soler, Phys. Status Solidi. B {\bf 215}, 809 (1999).

\bibitem{McCann}
E. McCann, K. Kechedzhi, V.I. Fal'ko, H. Suzuura, T. Ando, and B.L. Altshuler, Phys. Rev. Lett. {\bf 97}, 146805 (2006).

\bibitem{Mucha}
M. Mucha-Kruczy\'nski, O. Tsyplyatyev, A. Grishin, E. McCann, V.I. Fal'ko, A. Bostwick, and E. Rotenberg, Phys. Rev. B {\bf 77}, 195403 (2008).

\bibitem{Zhou}
S.Y. Zhou, D.A. Siegel, A.V. Fedorov, and A. Lanzara, Physica E {\bf 40}, 2642 (2008). 

\bibitem{Castro-Neto_RMP}
A.H. Castro Neto, F. Guinea, N.M.R. Peres, K.S. Novoselov, and A.K. Geim, Rev. Mod. Phys. {\bf 81}, 10 (2009).

\bibitem{Siesta-to-TB}
S. Dubois, A. Lherbier, A. Botello-Mendez, J.-C. Charlier, in preparation.

\bibitem{PRLKubo2011}
A. Lherbier, S.M.-M. Dubois, X. Declerck, S. Roche, Y.M. Niquet, and J.-C. Charlier, Phys. Rev. Lett. {\bf 106}, 046803 (2011).

\bibitem{Haydock}
R. Haydock, V. Heine, and M.J. Kelly, J. Phys. C : Solid State Phys. {\bf 8}, 2591 (1975).

\bibitem{broadeningdos}
The same broadening parameter ($10$ meV) has been used in both DOS calculations.

\bibitem{Wehling}
T.O. Wehling, S. Yuan, A.I. Lichtenstein, A.K. Geim, and M.I. Katsnelson, Phys. Rev. Lett. {\bf 105}, 056802 (2010).

\bibitem{Mesaros}
A. Mesaros, S. Papanikolaou, C.F.J. Flipse, D. Sadri, and J. Zaanen, Phys. Rev. B {\bf 82}, 205119 (2010).

\bibitem{GTrambly}
G. Trambly de Laissardi\`{e}re, and D. Mayou, Modern Phys. Lett. B {\bf 25}, 1019 (2011).

\bibitem{Lherbier-amorphous}
A. Lherbier, S. Roche, O.A. Restrepo, Y.M. Niquet, A. Decorte, and J.-C. Charlier, submitted for publication (2011).

\bibitem{Kapko}
V. Kapko, D.A. Drabold, and M.F. Thorpe, Phys. Status Solidi B {\bf 247}, 1197 (2010).

\bibitem{Holmstrom}
E. Holmstr\"om, J. Fransson, O. Eriksson, R. Liz\'arraga, B. Sanyal, S. Bhandary, and M.I. Katsnelson, Phys. Rev. B {\bf 84}, 205414 (2011).

\bibitem{Roche97}
S. Roche, and D. Mayou, Phys. Rev. Lett. {\bf 79}, 2518 (1997).

\bibitem{Mayou95}
D. Mayou, and S. Khanna, J. Phys. I France {\bf 5}, 1199 (1995).

\bibitem{Mayou2007}
D. Mayou, Europhys. Lett. {\bf 6}, 549 (2007).

\bibitem{Triozon2002}
F. Triozon, J. Vidal, R. Mosseri, and D. Mayou, Phys. Rev. B {\bf 65}, 220202 (2002).

\bibitem{Roche99}
S. Roche, Phys. Rev. B {\bf 59}, 2284 (1999).

\bibitem{Ishii}
H. Ishii, F. Triozon, N. Kobayashi, K. Hirose, S. Roche, C.R. Physique {\bf 10}, 283 (2009).

\bibitem{LeeRamakrishnan}
P.A. Lee, and T.V. Ramakrishnan, Rev. Mod. Phys. {\bf 57}, 287 (1985).

\bibitem{LeeFisher}
P.A. Lee, and D.S. Fisher, Phys. Rev. Lett. {\bf 47}, 882 (1981).

\bibitem{EversMirlin}
F. Evers and A.D. Mirlin, Rev. Mod. Phys. {\bf 80}, 1355 (2008).

\bibitem{Triozon2000}
F. Triozon, S. Roche, and D. Mayou, RIKEN Rev. {\bf 29}, 73 (2000).

\bibitem{note}
This also explain why the use of the same value of $D_{\text{max}}$ has been preferred to normalize the two equations in order to clearly see this spurious effect.

\bibitem{Shon}
N.H. Shon, and T. Ando, J. Phys. Soc. Jpn. {\bf 67}, 2421 (1998).

\bibitem{Ando}
T. Ando, International Journal of Modern Physics B {\bf 21}, 1113 (2007).

\bibitem{Roche2011}
S. Roche, B. Biel, A. Cresti, F. Triozon, Physica E (in press), DOI: 10.1016/j-physe.2011.06008 (2011).

\bibitem{Stauber}
T. Stauber, N.M.R. Peres, and F. Guinea, Phys. Rev. B {\bf 76}, 205423 (2007).

\bibitem{Yan2011}
J. Yan, and M.S. Fuhrer, Phys. Rev. Lett. {\bf 107}, 206601 (2011).

\bibitem{JHChen2008}
J.H. Chen, C. Jang, S. Xia, M. Ishigami, and M.S. Fuhrer, Nature Nanotech. {\bf 3}, 206 (2008).

\bibitem{Morozov}
S.V. Morozov et al. , Phys. Rev. Lett. {\bf 100}, 016602 (2008).

\bibitem{Montambaux_Akkermans}
E. Akkermans, and G. Montambaux, \textit{Mesoscopic Physics of Electrons and Photons}, Cambridge University Press, ISBN 978-0521855129 (2007).

\bibitem{Bang10}
J. Bang, and K.J. Chang, Phys. Rev. B {\bf 81}, 193412 (2010).

\bibitem{Skrypnyk}
Y.V. Skrypnyk, and V.M. Loktev, Phys. Rev. B {\bf 83}, 085421 (2011).

\bibitem{Schubert}
G. Schubert, and H. Fehske, arXiv:1109.6439v1 (2011).

\bibitem{Grassi}
R. Grassi, T. Low, M. Lundstrom, Nano Lett. {\bf 11}, 4574 (2011).

\bibitem{Elias}
D.C. Elias, R.R. Nair, T.M.G. Mohiuddin, S.V. Morozov, P. Blake, M.P Halsall, A.C. Ferrari, D.W. Boukhvalov, M.I. Katsnelson, A.K. Geim, and K.S. Novoselov, Science {\bf 323}, 610 (2009).

\bibitem{Bostwick}
A. Bostwick, J.L. McChesney, K.V. Emtsev, T. Seyller, K. Horn, S.D. Kevan, and E. Rotenberg, Phys. Rev. Lett. {\bf 103}, 056404 (2009).

\bibitem{LeconteACS}
N. Leconte, J. Moser, P. Ordej\'{o}n, H. Tao, A. Lherbier, A. Bachtold, F. Alsina, C.M. Sotomayor Torres, J.-C. Charlier, and S. Roche, ACS NANO {\bf 4}, 4033 (2010).

\bibitem{Moser10}
J. Moser, H. Tao, S. Roche, F. Alsina, C.M. Sotomayor Torres, and A. Bachtold, Phys. Rev. B {\bf 81}, 205445 (2010).

\bibitem{LecontePRB}
N. Leconte, A. Lherbier, F. Varchon, P. Ordej\'{o}n, S. Roche, and J.-C. Charlier, Phys Rev. B {\bf 84}, 235420 (2011).

\bibitem{Uppstu}
A. Uppstu, K. Saloriutta, A. Harju, M. Puska, and A.-P. Jauho, arXiv:1111.0148v1 (2011), submitted to Phys. Rev. Lett.

\bibitem{Haskins}
J. Haskins, A. Kinaci, C. Sevik, H. Sevin\c{c}li, G. Cuniberti, and T. \c{C}a\u{g}in, ACS NANO {\bf 5}, 3779 (2011).

\bibitem{Wu}
S. Wu, and F. Liu, arXiv:1001.2057v1 (2010).

\bibitem{WeiLi11}
W. Li, Y. He, L. Wang, G. Ding, Z-.Q. Zhang, R. W. Lortz, P. Sheng, and N. Wang, Phys. Rev. B {\bf 84}, 045431 (2011).

\end{thebibliography}
\end{document}